\documentclass[11pt,a4paper]{article}
\usepackage[T1]{fontenc}
\usepackage[utf8]{inputenc}
\usepackage{lmodern}
\usepackage[a4paper,margin=25mm]{geometry}
\usepackage{amsmath,amssymb,bm}
\usepackage{graphicx}
\usepackage{booktabs,tabularx,array}
\usepackage[font=small,labelfont=bf]{caption}
\usepackage[numbers,sort&compress]{natbib}
\usepackage{microtype}
\usepackage{xcolor}
\usepackage[colorlinks=true,linkcolor=blue!45!black,citecolor=blue!45!black,urlcolor=blue!45!black]{hyperref}
\hypersetup{pdftitle={Task alignment and configuration selection in SNAP quantum reservoir computing}}
\newcounter{lem}
\newcounter{thm}
\newcounter{cor}
\newcommand{\manuscriptfigure}[2][\linewidth]{%
  \IfFileExists{figures/figure#2.pdf}{\includegraphics[width=#1]{figures/figure#2.pdf}}{%
    \includegraphics[width=#1]{figures/figure#2.png}}}

\title{Task-aligned bosonic quantum reservoir computing\\ with SNAP encoding}

\author{%
  Lu-Fan Zhang$^{1}$,
  Yusen Wu$^{1}$,
  Yong-Pan Gao$^{2}$,
  Chuan Wang$^{1,3}$%
  \thanks{Chuan Wang: wangchuan@bnu.edu.cn}\\[0.6em]
  {\small $^{1}$School of Artificial Intelligence, Beijing Normal University, 100875, Beijing, China}\\
  {\small $^{2}$School of Electronics Engineering,} \\ {\small Beijing University of Posts and Telecommunications, 100876, Beijing, China}\\
  {\small $^{3}$Engineering Research Center of Intelligent Technology and Educational Application,} \\ {\small Beijing Normal University, 100875, Beijing, China}\\
  {\small $^{*}$Author to whom any correspondence should be addressed.}
}

\date{}

\begin{document}
\maketitle

\begin{abstract}
\noindent Quantum reservoir computing (QRC) maps input histories to measurable features that a classical readout uses for prediction. Effective task-specific reservoir design requires understanding which history-dependent functions are accessible to the readout and how well they align with the target in addition to aggregate metrics such as information processing capacity. Here, we develop a task-alignment framework for bosonic QRC in which selective number-dependent arbitrary phase (SNAP) encoding generates nonlinear input responses, hopping propagates them through time, and first and second moments serve as readout features. Analytically, we characterize the encoding-induced Fourier function space, derive the input-dependent moment functions and their delay-propagation recursions, and establish a finite-history fading-memory bound. To express the target in terms of nonlinear, delayed, and interacting history functions, we introduce a Legendre teacher model. Nested projections then decompose the prediction risk into candidate-function, alignment, and readout contributions, yielding a configuration-selection criterion with accuracy guarantees. Experiments on the Mackey-Glass, ETT, and Weather datasets show that task-aligned configurations improve forecasting accuracy and match or exceed state-of-the-art methods across several benchmarks.
\end{abstract}

\providecommand{\keywords}[1]{%
  \begin{quote}\small\noindent\textbf{Keywords:} #1\end{quote}%
}


\section{Introduction}

Quantum algorithms for integer factorization and unstructured search demonstrated how quantum resources can improve computational scaling for specific problems \cite{Shor1994Algorithms,Grover1996DatabaseSearch}. These advances helped motivate quantum machine learning (QML), which explores how quantum states and operations can support learning from data \cite{Biamonte2017QuantumMachineLearning,Cerezo2022ChallengesQML,RodriguezDiaz2025QMLSurvey}. Two prominent approaches are parameterized quantum circuits and quantum feature spaces \cite{Mitarai2018QuantumCircuitLearning,Schuld2019FeatureHilbertSpaces,Havlicek2019QuantumFeatureSpaces}. Applications have been explored in multimodal diagnosis, healthcare, and drug discovery \cite{Qu2023QNMF,Aggarwal2023QuantumHealthcare,Smaldone2025DrugDiscovery}, while theoretical work has investigated the conditions under which quantum learning can offer computational advantages \cite{Yamasaki2026QMLAdvantage}. Within this landscape, quantum reservoir computing (QRC) uses a random fixed quantum dynamics to map inputs to observable features, confining training to a classical readout \cite{Fujii2017QuantumReservoir,Tanaka2019PhysicalReservoirReview}. For quantum tasks, reservoir-based processing has been applied to circuit compression, state-property estimation, quantum tomography, entanglement witnessing, and probing quantum phase transitions \cite{Ghosh2019QuantumReservoirProcessing,Ghosh2021QuantumCircuitCompression,Li2024QuantumStateProperties,Krisnanda2025QuantumTomography,Zia2025EntanglementWitnessing,Kobayashi2025ReservoirProbing}. On classical data, applications span microwave signal classification, image classification in spin and photonic systems, and molecular property prediction \cite{Senanian2024MicrowaveReservoir,Zhang2026DiscreteTimeCrystal,Cimini2026GaussianBosonSampler,Beaulieu2025MolecularPropertyPrediction}, as well as forecasting temporal signals and chaotic dynamics \cite{Hou2026TemporalPrediction,Connerty2026ChaoticDynamics}. Across these applications, computational performance hinges on how closely the information accessible to the readout reflects the structure of the target.

For time-series forecasting, this structure involves nonlinear dependence on past inputs, memory at specific delays, and interactions between inputs at different times. Short-term memory measures and information processing capacity (IPC) characterize these capabilities and have been studied experimentally in classical reservoirs and theoretically in spin-based QRC \cite{Dambre2012InformationProcessingCapacity,Vettelschoss2022SingleNodeIPC,MartinezPena2023SpinReservoirIPC}. Relating such aggregate measures to performance on a specific target, however, requires examining how capacity is distributed across task-relevant functions. Task-specific IPC analysis formalizes this link by connecting prediction error directly to the capacity allocated to the required functions \cite{Hulser2023TaskSpecificIPC}. Studies of quantum reservoirs have also examined memory restriction and the distinguishability of input histories \cite{Cindrak2024ArtificialMemoryRestriction,MartinezPena2025InputDependence}. Other work has explored how memory depends on optical response, non-Markovian dynamics, many-body regimes, and reservoir architecture \cite{Gotting2025MemoryOpticalAbsorption,Sannia2026NonMarkovianMemory,Kobayashi2026QuantumChaos,Benavoli2026QuantumWiener}. Complementary Fourier analyses characterize constraints on the history-dependent functions generated by quantum encodings, while Krylov-based studies connect expressivity and observability to computational performance \cite{Schutte2026Expressivity,Cindrak2025KrylovObservability}. Finite sampling further determines how reliably the available computational directions can be resolved \cite{Hu2023Eigentasks}. These studies motivate a task-specific analysis of the history-dependent functions accessible through a physical reservoir. Under squared-error loss, the optimal predictor based on the input history is the conditional expectation of the target. At the population level, the optimal unregularized linear readout is the projection of this conditional expectation onto the span of the measured reservoir features, with the projection defined by the task's input distribution. This formulation relates prediction error to the alignment between the target and the readable history-function space. Characterizing this space requires understanding how encoding, memory propagation, and the choice of observables jointly determine which dependencies on past inputs are accessible.

Here we introduce a bosonic QRC architecture that combines selective number-dependent arbitrary phase (SNAP) input encoding with number-conserving hopping interactions for memory propagation. SNAP operations provide photon-number-dependent phase control and have been demonstrated experimentally, with subsequent advances in gate design \cite{Heeres2015SNAP,You2025FastSNAP,Meschede2026EigenSNAP}. Previous studies of bosonic QRC have explored Gaussian oscillator networks, optical continuous-variable systems, circuit-QED devices, and Jaynes--Cummings models \cite{Nokkala2021GaussianReservoir,Paparelle2026OpticalMemoryControl,Carles2026CircuitQEDReservoir,Das2026JaynesCummingsReservoir}. In our architecture, eigenvalue differences of the SNAP encoding generator constrain the candidate Fourier components of the input response, consistent with analyses of quantum data encoding and encoding-induced nonlinearities \cite{Schuld2021DataEncoding,Govia2022NonlinearInput}. The reference state and the choice of first and second moments determine which components enter the memory dynamics. Hopping interactions then distribute these responses across memory modes and temporal delays, and the measured observables determine which history-dependent functions remain accessible to a linear readout. This gives three levels of description: the frequencies present in the encoded state, the functions carried by the accessible moments, and the linearly independent history functions available to the readout. We derive the encoded moment responses and their propagation to establish how these physical constraints shape the readable history-function space.

Building on this description, we relate forecasting error to the alignment between the predictable component of the target and the history-dependent functions accessible to the SNAP reservoir's readout. Our analysis identifies the functions supported by a given encoding, finite memory, and set of observables, and assesses how well they represent the target. We evaluate this framework by varying encoding and propagation parameters separately and comparing task-aligned configurations with those selected using conventional memory metrics. We then assess forecasting accuracy at multiple prediction horizons on the Mackey--Glass system and the ETT~\cite{Zhou2021Informer} and Weather datasets~\cite{MPIBGCWeather}, covering both single-channel and multichannel prediction and comparisons with established forecasting methods. Taken together, our analysis and experiments establish function--target alignment as a principled criterion for selecting SNAP reservoir configurations in forecasting tasks.

\section{Forecasting requirements and readable history functions}\label{sec:1}

For time-series forecasting, the measured reservoir features need to encode historical information in a form accessible to the readout. Thus we formulate the design of a quantum reservoir around three task requirements, including nonlinear functions of individual inputs, information at relevant delays, and interactions between inputs at different times. We derive the history-dependent functions generated by SNAP encoding, passive hopping, and moment observations. A Legendre teacher then represents the target in terms of these requirements, and its projection onto the readable function space guides configuration selection (Sec.~\ref{sec:2}). Finally, error decompositions relate this task-aligned selection criterion to physical resources, teacher accuracy, and the performance of the fitted predictor.

\subsection{Task requirements and capacity indicators}
We begin with the time-series forecasting task. Let $u_{t}$ be a scalar input at discrete time $t$ and $\mathbf{u}_{t} = ( u_{t} , u_{t - 1} , \ldots )$ its input history vector. Any function of $\mathbf{u}_{t}$ is a \emph{history function}. The target $Y_{t}$ is the future time-series value to be predicted from history $\mathbf{u}_{t}$. The expectation values of selected reservoir observables serve as readout features. Including a constant coordinate $1$, we collect them into the real feature vector $X_{t} = X ( \mathbf{u}_{t} ) = ( 1 , X_{1} ( \mathbf{u}_{t} ) , \ldots , X_{n_{X} - 1} ( \mathbf{u}_{t} ) )^{\top}$ with length $n_{X}$. A linear readout with a bias gives the prediction
\begin{equation}
\widehat Y_t
= w_0+\sum_{j=1}^{n_X-1}w_jX_j(\mathbf{u}_t)
= w^\top X_t ,
\end{equation}
where $w_{0}$ is the bias, $w_{j}$ weights the jth readout feature, and $w = ( w_{0} , w_{1} , \ldots , w_{n_{X} - 1} )^{\top}$ is the real readout weight vector. Each readout feature \(X_j\) is a real-valued function of the input history. We use the real and imaginary parts of complex moments as separate features. These readout features linearly span a \emph{readable function space} $\mathcal{S}$, which contains all prediction functions obtainable by varying the readout weights at a fixed reservoir configuration. The encoding, propagation, and observation choices are held fixed when defining this space. Training the readout selects a function in $\mathcal{S}$ to fit the target $Y_{t}$ under the chosen fitting criterion.

We consider three requirements that can arise in time-series forecasting. First, the measured reservoir features should allow the linear readout to approximate nonlinear functions of individual inputs. Second, information about past inputs should remain accessible to the readout at the delays relevant to the task. Third, the measured features should capture interactions between inputs at different times. For example, the term \(u_t u_{t-k}^2\) combines all three requirements. It depends quadratically on the input from \(k\) time steps earlier and couples that input to the current one. We compare the history functions required by the task with those accessible through the reservoir readout. For a given task, all projections use the same joint distribution \(P\) of input histories and targets. We assume that the target and all feature functions considered below have finite second moments under \(P\). Each function \(s\in\mathcal S\) gives a prediction \(s(\mathbf u_t)\) for the input history \(\mathbf u_t\). The optimal readout \(s^\star\) minimizes the mean squared prediction error under \(P\),
\begin{equation}
s^\star=\arg\min_{s\in\mathcal S}\mathbb E_P\!\left[(Y_t-s(\mathbf u_t))^2\right].
\end{equation}

To test which history functions the reservoir can reproduce, we use a specified function \(e\in L^2(P)\) as the target of the linear readout. For a centered test function with unit variance, the information processing capacity (IPC) is the fraction of the variance recovered by the optimal readout \citep{Dambre2012InformationProcessingCapacity},
\begin{equation}
C(e)=1-\min_{s\in\mathcal S}\mathbb E_P[(e-s)^2]=\|\Pi_{\mathcal S}e\|_2^2,
\end{equation}
where \(\Pi_{\mathcal S}e\) is the optimal mean-square approximation to \(e\). \(C(e) = 1\) indicates exact recovery, while \(C(e) = 0\) means that the readout captures none of the variation in \(e\).

For the IPC tests considered here, inputs are independent and uniformly distributed on \([-1,1]\). Products of normalized Legendre polynomials then form an orthonormal basis for square-integrable functions of a finite input history:
\[
e_{\boldsymbol n}(\mathbf u_{t,L})=\prod_{k=0}^{L-1}\sqrt{2n_k+1}\,P_{n_k}(u_{t-k}),
\]
where \(P_{n_k}\) is the Legendre polynomial of degree \(n_k\). This basis allows the three forecasting requirements to be evaluated separately. At a fixed delay, a polynomial of degree \(n\geq 2\) tests whether the readout can recover the corresponding nonlinear function of a single input. A degree-one polynomial at delay \(k\) tests linear recall of \(u_{t-k}\), and summing these capacities over the tested past delays gives the linear short-term memory (STM) capacity. Products with nonconstant factors at two or more distinct delays test interactions between inputs at different times. The total IPC sums the capacities of the tested functions.

The predictable component of a forecasting target is a task-specific combination of history functions. In Fig.~\ref{fig:1}(b), two reservoirs with the same total capacity yield different prediction errors because their readable spaces capture different components of the target. For a task with input distribution \(P\), we quantify task alignment through the minimum prediction risk \(R_*\), whose configuration-dependent contribution is the squared \(L^2(P)\) distance from the predictable component of the target to the readable space \(\mathcal S\):
\begin{equation}
R_* = \min_{g\in\mathcal{S}}\mathbb{E}_P[(Y-g)^2]
      =\sigma_{\mathrm{irr}}^2+\|f-\Pi_{\mathcal{S}}f\|_2^2,
\end{equation}
where $f(\mathbf{u}) = \mathbb{E}_{P} [Y\mid\mathbf{u}]$ is the predictable component of the target and $\sigma_{\mathrm{irr}}^{2} = \| Y - f \|_{2}^{2}$ is the irreducible error. For a fixed task and input distribution, $\sigma_{\mathrm{irr}}^{2}$ is constant across reservoir configurations. A smaller $R_*$ indicates better task alignment, with the lower bound $R_*=\sigma_{\mathrm{irr}}^2$ attained when $f\in\mathcal S$. This criterion guides the choice of encoding, memory dynamics, and observables that make the required history functions accessible within the available physical resources.

\subsection{History functions generated by SNAP QRC}\label{sec:history-functions}

\begin{figure}[tbp]
\centering
\manuscriptfigure{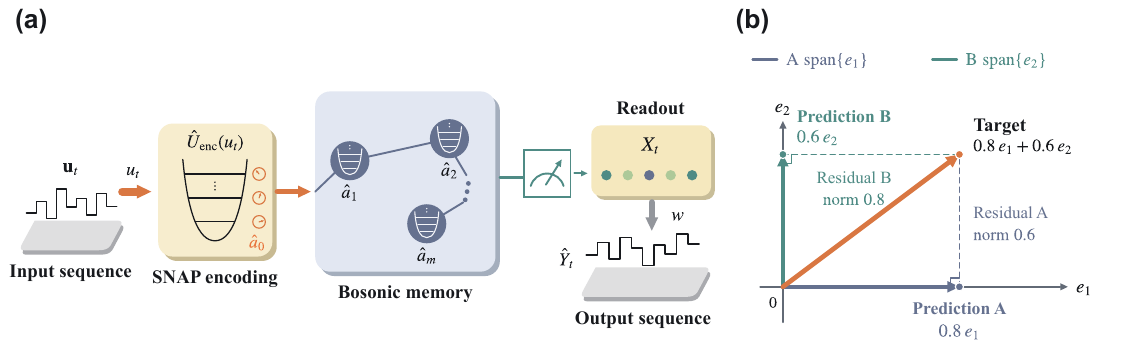}
\caption{\textbf{SNAP QRC architecture and task alignment.} (a) The reservoir consists of one encoding mode $\hat a_0$ and $m$ persistent memory modes $\hat a_j$. Each network node represents a bosonic mode modeled as a quantum harmonic oscillator. For mode $j$, each rung corresponds to a Fock state $\lvert n\rangle_j$ with $n$ excitations. The SNAP unitary $\hat U_{\mathrm{enc}}(u_t)$ applies input-dependent phases to the Fock states of the encoding mode. The encoding mode is reset before each input. Hopping transfers the encoded input to the memory modes. The linear readout combines the measured features to produce the forecast $\widehat Y_t=w^\top X_t$. (b) The centered readable spaces of configurations $A$ and $B$ are spanned by $e_1$ and $e_2$, respectively, where $e_1$ and $e_2$ are orthonormal under the same input distribution. Both spaces have dimension one and summed capacity one over these two test functions. For the target $0.8e_1+0.6e_2$, the dashed segments show the projection residuals, $0.6e_2$ for $A$ and $0.8e_1$ for $B$. Their squared norms give the corresponding mean squared errors of $0.36$ and $0.64$.}
\label{fig:1}
\end{figure}

Figure~\ref{fig:1}(b) illustrates the effect of task alignment for two prescribed readable spaces. We now derive the history functions that span the readable space of the SNAP QRC architecture in Fig.~\ref{fig:1}(a). For a scalar input encoded by \(e^{iuG}\), the eigenvalue gaps of the generator \(G\) specify which Fourier frequencies can appear in the response to \(u\) \citep{Schuld2021DataEncoding}. In a reservoir with input-independent propagation, this constraint applies separately to each input in the history \citep{Schutte2026Expressivity}. We choose a SNAP gate because it allows the phase of each Fock level to be set independently \citep{Heeres2015SNAP}. By varying the rate at which each phase changes with \(u\), we can tune the nonlinear input functions carried by the first and second moments while keeping the hopping dynamics fixed.

At each time step, a SNAP gate encodes the current input in a freshly prepared bosonic mode. Hopping couples this mode to \(m\) persistent memory modes. The input mode is then discarded, and the memory state is retained for the next step. The linear readout predicts the target from the real and imaginary parts of the first and second moments of the memory modes. To identify the input dependence that enters these moments, we begin with the action of the SNAP gate on a single input \(u\). On the occupied Fock levels, the encoding unitary reads
\begin{equation}
\hat U_{\mathrm{enc}}(u)
= e^{iu\hat G_{\mathrm{enc}}}
= \sum_{n=0}^{d_{\mathrm F}-1}e^{i\lambda_nu}\lvert n\rangle\langle n\rvert,
\end{equation}
where \(\lvert n\rangle\) is the Fock state with \(n\) excitations, and \(d_{\mathrm F}\) is the number of levels in the chosen input support. The encoding generator \(\hat G_{\mathrm{enc}}=\sum_{n=0}^{d_{\mathrm F}-1}\lambda_n\lvert n\rangle\langle n\rvert\) assigns the phase \(\lambda_n u\) to level \(n\). At each step the input mode is prepared in the same reference state \(\lvert\psi_0\rangle=\sum_{n=0}^{d_{\mathrm F}-1}c_n\lvert n\rangle\), with \(\sum_n|c_n|^2=1\). After encoding, the state and its density operator are
\begin{equation}
\lvert\psi(u)\rangle=\sum_{n=0}^{d_{\mathrm F}-1}c_ne^{i\lambda_nu}\lvert n\rangle,
 \qquad
 \hat{\rho}_{\mathrm{in}}(u)=\sum_{n,j=0}^{d_{\mathrm F}-1}
 c_nc_j^*e^{i(\lambda_n-\lambda_j)u}\lvert n\rangle\langle j\rvert.
\label{eq:3}
\end{equation}

Here, the indices \(n\) and \(j\) label Fock levels, and the star denotes complex conjugation. For a state \(\hat\rho\), the expectation value of an operator \(\hat O\) is denoted by \(\langle\hat O\rangle=\operatorname{Tr}(\hat\rho\hat O)\). A complex moment is reconstructed from the expectation values of the Hermitian real and imaginary parts of the corresponding operator.

To vary the shape of the encoding spectrum at fixed spectral standard deviation \(s_G\), we set \(\lambda_n(\beta)=s_G(\beta^n-\bar b_\beta)/\sigma_\beta\) for \(\beta>1\), where \(\bar b_\beta=d_{\mathrm F}^{-1}\sum_n\beta^n\) and \(\sigma_\beta^2=d_{\mathrm F}^{-1}\sum_n(\beta^n-\bar b_\beta)^2\). The limit \(\beta\to1^+\) recovers an equally spaced spectrum. In Eq.~\eqref{eq:3}, each coherence \(c_nc_j^*\) contributes at frequency \(\lambda_n(\beta)-\lambda_j(\beta)\). The frequencies present in the encoded state form the set \(\Omega_{\mathrm{state}}(\beta)=\{\lambda_n(\beta)-\lambda_j(\beta)\mid c_nc_j^*\ne0\}\). We define the \emph{encoding Fourier space} as the real span of the constant function and the functions \(\cos(\omega u)\) and \(\sin(\omega u)\) for all positive \(\omega\in\Omega_{\mathrm{state}}(\beta)\). If there are \(K_\Omega(\beta)\) such frequencies, this space has dimension \(1+2K_\Omega(\beta)\leq1+d_{\mathrm F}(d_{\mathrm F}-1)\) on any input interval of nonzero length. Measuring arbitrary Hermitian observables on the full encoded state provides access to this entire space. In contrast, the moments that drive the memory select specific combinations of these frequencies.

\refstepcounter{lem}\label{lem:1}
\noindent\textbf{Lemma~\thelem. Input moment functions at fixed \(\beta\).} Let \(\hat a_0\) be the annihilation operator of the input mode. For an integer \(1\leq p<d_{\mathrm F}\), the encoded input has moment
\begin{equation}
\mu_p(u,\beta)=\langle \hat{a}_0^p\rangle
 =\sum_{n=0}^{d_{\mathrm F}-p-1}\sqrt{\frac{(n+p)!}{n!}}\,
 c_n^*c_{n+p}e^{i[\lambda_{n+p}(\beta)-\lambda_n(\beta)]u} .
\label{eq:4}
\end{equation}

Equation~\eqref{eq:4} follows by applying \(\hat a_0^p\) to the Fock states in Eq.~\eqref{eq:3}. In particular, \(\mu_1\) samples coherences between adjacent Fock levels, and \(\mu_2\) samples coherences between levels separated by two excitations. The photon-number moment \(\nu=\langle\hat a_0^\dagger\hat a_0\rangle=\sum_n n|c_n|^2\) is constant because SNAP encoding preserves the Fock populations. The input dependence of the first and second moments is thus described by four real \emph{input moment functions},
\begin{equation}
\begin{aligned}
\xi_1(u,\beta)&=\operatorname{Re}\mu_1(u,\beta), &\xi_2(u,\beta)&=\operatorname{Im}\mu_1(u,\beta),\\
\xi_3(u,\beta)&=\operatorname{Re}\mu_2(u,\beta), &\xi_4(u,\beta)&=\operatorname{Im}\mu_2(u,\beta).
\end{aligned}
\end{equation}
At fixed \(\beta\), we count the independent nonconstant combinations of these functions separately from the readout intercept. Their rank modulo constants is
\begin{equation}
r_{\mathrm{src}}(\beta)
=\dim\operatorname{span}_{\mathbb R}
\{1,\xi_1(\cdot,\beta),\ldots,\xi_4(\cdot,\beta)\}-1
\leq 4 .
\end{equation}
For a fixed encoding, we write \(\mu(u)=\mu_1(u,\beta)\) and \(\zeta(u)=\mu_2(u,\beta)\). Each is a weighted sum of Fourier components, with weights fixed by the reference state and the matrix elements of the corresponding operator in the Fock basis. Changing \(\beta\) shifts the frequencies in these sums and can therefore alter the input functions even when \(r_{\mathrm{src}}\) remains unchanged (Fig.~\ref{fig:2}(a,b)).

In the numerical experiments, we use \(d_{\mathrm F}=8\), \(c_n=1/\sqrt{8}\), and \(s_G=11\sqrt{21}/40\), giving \(\nu=7/2\). The eight-level support applies only to input preparation, whereas the memory evolves in the full bosonic mode space. With these parameters, the encoding Fourier space has dimension \(15\) in the limit \(\beta\to1^+\) and \(57\) at both \(\beta=1.02\) and \(\beta=3\), counting the constant function. At both \(\beta=1.02\) and \(\beta=3\), \(\mu\) contains seven distinct positive frequencies and \(\zeta\) contains six. The nonconstant input moment rank is four in all three cases. Comparing \(\beta=1.02\) and \(\beta=3\) therefore isolates a change in function shape at fixed Fourier dimension and input moment rank. Appendix~\ref{app:A2} gives the derivation. These dimensions reflect exact linear independence on the input interval. With sampled inputs, closely spaced frequencies may be difficult to resolve, and the effective dimension depends on normalization and on the threshold used to retain singular directions.

The functions \(\mu\) and \(\zeta\) describe a single encoded input. Their contributions at later times depend on propagation through the memory. We couple the fresh input mode to the memory through the number-conserving hopping Hamiltonian
\begin{equation}
\hat H=\sum_{j=0}^{m-1}g_j
\left(\hat a_j^\dagger\hat a_{j+1}
+\hat a_{j+1}^\dagger\hat a_j\right).
\end{equation}
Here, \(g_j\) is a real coupling and \(\hat a_j\) is the annihilation operator for mode \(j\). Mode \(0\) encodes the input, and modes \(1,\ldots,m\) retain the memory. In units with \(\hbar=1\), an interaction of duration \(\tau\) generates the unitary \(\hat U=\exp(-i\tau\hat H)\). Tracing out the input mode after the interaction gives the memory update
\begin{equation}
\hat\rho_t^{(M)}
=\operatorname{Tr}_0\!\left[
\hat U\left(\hat\rho_{\mathrm{in}}(u_t)\otimes
\hat\rho_{t-1}^{(M)}\right)\hat U^\dagger
\right],
\end{equation}
where \(\operatorname{Tr}_0\) denotes the partial trace over mode \(0\). We parameterize the interaction time by \(\theta=g_{\mathrm{ref}}\tau\), where \(g_{\mathrm{ref}}\) is the nominal coupling scale. We index the memory immediately after propagation, so the current input \(u_t\) has age zero.

To track an input over successive steps, let \(\mathsf h\) denote the single-particle Hamiltonian matrix and partition its propagator into input and memory blocks,
\begin{equation}
e^{-i\tau\mathsf h}
=\begin{pmatrix}s&\ell\\ b&A\end{pmatrix}.
\end{equation}
The corresponding update for the retained annihilation operators is
\begin{equation}
\hat{\boldsymbol r}_t
=A\hat{\boldsymbol r}_{t-1}+b\hat a_{0,t}.
\label{eq:5}
\end{equation}
In Eq.~\eqref{eq:5}, \(\hat{\boldsymbol r}_t\) collects the \(m\) memory annihilation operators, and \(\hat a_{0,t}\) acts on the fresh input. The \(m\times m\) matrix \(A\) propagates the memory, and the column vector \(b\) transfers the input into it. The row vector \(\ell\) transfers memory into the discarded port, and \(s\) is the amplitude for the input to remain in that port. Conditional on the classical input history, the fresh input and the previous memory are in a product state. This factorization closes the moment equations at second order even when the classical inputs are correlated.

We collect the observable moments in the mean vector \(\alpha_t\), the Hermitian normal-moment matrix \(Q_t\), and the symmetric anomalous-moment matrix \(F_t\). For propagation, it is convenient to introduce their centered second moments \(V_t\) and \(W_t\),
\begin{equation}
\begin{aligned}
\alpha_t&=\langle\hat{\boldsymbol r}_t\rangle,\\
[Q_t]_{ij}&=\langle\hat r_{t,j}^\dagger\hat r_{t,i}\rangle,\\
[F_t]_{ij}&=\langle\hat r_{t,i}\hat r_{t,j}\rangle,\\
V_t&=Q_t-\alpha_t\alpha_t^\dagger,\\
W_t&=F_t-\alpha_t\alpha_t^\top.
\end{aligned}
\end{equation}
Indices \(i\) and \(j\) label memory modes, and \(\dagger\) and \(\top\) denote the adjoint and transpose, respectively. Subtracting the products of means in \(V_t\) and \(W_t\) gives simple recursions. Applying Eq.~\eqref{eq:5} to the conditional product state yields
\begin{equation}
\begin{aligned}
\alpha_t&=A\alpha_{t-1}+b\mu(u_t),\\
V_t&=AV_{t-1}A^\dagger+\kappa(u_t)bb^\dagger,\\
W_t&=AW_{t-1}A^\top+\chi(u_t)bb^\top,
\end{aligned}
\label{eq:6}
\end{equation}
with input terms \(\kappa(u)=\nu-|\mu(u)|^2\) and \(\chi(u)=\zeta(u)-\mu(u)^2\). An input injected through \(b\) and propagated for \(k\) further steps has response \(v_k=A^kb\) for \(k\geq0\) (Fig.~\ref{fig:2}(c)). Thus, \(v_k\) represents its amplitude in each memory mode at age \(k\). The two-sided matrix actions in Eq.~\eqref{eq:6} determine the corresponding responses for the second moments (see Appendices~\ref{app:A3} and~\ref{app:A4} for derivations).

We can now express the measured moments directly as functions of the input history. Assume that the spectral radius satisfies \(\operatorname{spr}(A)<1\), and write \(\mu^{(k)}=\mu(u_{t-k})\) and \(\zeta^{(k)}=\zeta(u_{t-k})\). Iterating Eq.~\eqref{eq:6} and restoring the products of means gives the stationary causal response
\begin{equation}
\begin{aligned}
\alpha_t&=\sum_{k\geq0}v_k\mu^{(k)},\\
Q_t&=\nu\sum_{k\geq0}v_kv_k^\dagger
+\sum_{\substack{k,j\geq0\\k\ne j}}v_kv_j^\dagger\mu^{(k)}(\mu^{(j)})^*,\\
F_t&=\sum_{k\geq0}v_kv_k^\top\zeta^{(k)}
+\sum_{\substack{k,j\geq0\\k\ne j}}v_kv_j^\top\mu^{(k)}\mu^{(j)} .
\end{aligned}
\label{eq:7}
\end{equation}
The double sums run over ordered pairs of distinct ages. For memory initialized in vacuum and driven by \(L\) inputs, all indices range from \(0\) to \(L-1\). Upon restoring the products of means, the equal-age terms \(|\mu^{(k)}|^2\) and \((\mu^{(k)})^2\) cancel. The remaining terms in \(Q_t\) consist of a constant and products across distinct ages. In \(F_t\), the single-age terms retain the second input moment \(\zeta^{(k)}\). For stable single-port propagation, \(\sum_{k\geq0}v_kv_k^\dagger=I_m\), so the constant in the stationary \(Q_t\) is \(\nu I_m\). Equation~\eqref{eq:7} identifies the origin of the three forecasting requirements. SNAP encoding supplies the nonlinear functions \(\mu\) and \(\zeta\) of individual inputs. The vectors \(v_k\) and their outer products weight the contributions at each delay. Products of first moments at distinct ages give the temporal interactions. A linear readout of the first and second moments selects real linear combinations of these terms, so every contributing term depends on at most two distinct input times.

For a window \(\mathbf u_{t,L}=(u_t,\ldots,u_{t-L+1})\), let \(z_L\) collect the \emph{candidate history functions} in Eq.~\eqref{eq:7}. Its entries are the constant \(1\), the single-input terms \(\xi_a(u_{t-k})\) for \(a=1,\ldots,4\) and \(0\leq k<L\), and the products \(\xi_a(u_{t-k})\xi_b(u_{t-j})\) for \(a,b\in\{1,2\}\) and \(0\leq k<j<L\). At fixed \(\beta\), their real span in \(L^2(P)\) is the candidate space \(\mathcal D_L\). The number of entries in \(z_L\) is
\begin{equation}
N_L=1+4L+4\binom{L}{2}.
\end{equation}
Linear dependence under \(P\) can reduce the dimension of \(\mathcal D_L\). Propagation and measurement impose a further constraint. By Eq.~\eqref{eq:7}, the feature vector has the form \(X_L=\mathsf B_Lz_L\), where the real matrix \(\mathsf B_L\) specifies the combinations accessible to the selected observables. Observing all first and second moments yields \(n_{\mathrm{obs}}=2m^2+3m\) real coordinates, comprising \(2m\) from the means, \(m^2\) from the normal moments, and \(m(m+1)\) from the anomalous moments. For \(m=6\), this amounts to \(90\) real coordinates. Let \(n_X\) count the selected coordinates and the intercept, and let \(\mathcal S_L\) denote their readable function space.

\refstepcounter{thm}\label{thm:1}
\noindent\textbf{Theorem~\thethm. Readable dimension and finite-window realization.} With \(\Gamma_L=\mathbb E_P[z_Lz_L^\top]\), the dimensions of the candidate and readable spaces satisfy
\begin{equation}
\begin{aligned}
\dim\mathcal D_L&=\operatorname{rank}\Gamma_L\leq N_L,\\
\mathcal S_L&\subseteq\mathcal D_L,\qquad
\dim\mathcal S_L=\operatorname{rank}(\mathsf B_L\Gamma_L^{1/2})
\leq\min\{n_X,n_{\mathrm{obs}}+1,\dim\mathcal D_L\}.
\end{aligned}
\end{equation}

The Gram matrix \(\Gamma_L\) accounts for dependencies among the candidate functions, and \(\mathsf B_L\) determines which combinations reach the readout. If the memory starts in vacuum, all first and second moments are observed, and \(\mathsf K_L=[b,Ab,\ldots,A^{L-1}b]\) has column rank \(L\), then \(\mathcal S_L=\mathcal D_L\). A left inverse of \(\mathsf K_L\) recovers the candidate functions as linear combinations of the measured moments (Appendix~\ref{app:A5}). For the stable single-port unitary model, this rank condition holds for \(L\leq m\). Thus, a vacuum-initialized memory with \(m\geq L\) modes and complete observations can realize the entire candidate space on that window. The conditioning of \(\mathsf K_L\) determines the sensitivity of the recovery. Even under exact recovery, each candidate term involves at most two distinct input times (Corollary~\ref{cor:1}).

\subsection{Memory responses and finite-history approximation}

To relate this finite-window result to continuous operation, we bound the contribution of older inputs.

\refstepcounter{lem}\label{lem:2}
\noindent\textbf{Lemma~\thelem. Leakage and history response.} Unitarity gives $A^{\dagger} A + \ell^{\dagger} \ell = I_{m}$ and $A A^{\dagger} + b b^{\dagger} = I_{m}$. Consequently, $\|v_{k+1}\|_{2}^{2} = \|v_{k}\|_{2}^{2} - |\ell v_{k}|^{2}$. For any $r$ satisfying $\operatorname{spr}(A) < r < 1$, there exists a finite constant $C_{A}$ such that $\|A^{k}\|_{2} \leq C_{A} r^{k}$ for all $k \geq 0$. The total response energy decays through leakage into the discarded port, while individual mode amplitudes may oscillate. This power bound also covers stable matrices with $\|A\|_{2} = 1$. Appendix~\ref{app:A3} proves the leakage, controllability, and response-sum identities.

\begin{figure}[tbp]
\centering
\manuscriptfigure{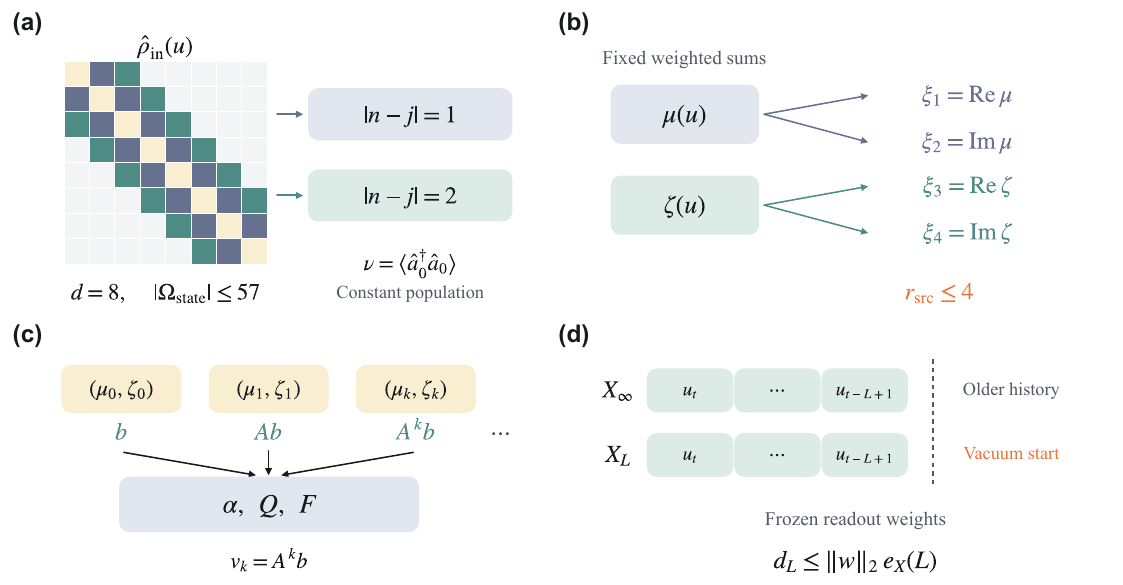}
\caption{\textbf{Encoding, propagation, and readable history functions.} (a) Adjacent and next-nearest Fock coherences supply $\mu = \mu_{1}$ and $\zeta = \mu_{2}$, whereas the number moment $\nu$ is input independent. (b) Their fixed weighted Fourier sums give four real input moment functions $\xi_{a}$ at each $\beta$, with nonconstant rank $r_{\mathrm{src}}(\beta) \leq 4$. (c) The response $v_{k} = A^{k} b$ assigns coefficients to input age $k$. Raw moments contain the single-time functions and two-time products in Eq.~\eqref{eq:7}. (d) Full-history features $X_{\infty}$ and vacuum-initialized features $X_{L}$ share the same $L$ recent inputs. With frozen readout weights, their output discrepancy satisfies $d_{L} \leq \|w\|_{2} e_{X}(L)$. The panels depict the analytic construction.}
\label{fig:2}
\end{figure}

\refstepcounter{thm}\label{thm:2}
\noindent\textbf{Theorem~\thethm. Moment fading memory.} For continuous input moment functions on a compact input interval with $\operatorname{spr}(A) < 1$, the moment dynamics admit a unique bounded causal response. For any fixed real linear observation and fixed affine normalization, the full-history features $X_{\infty}$ and vacuum-initialized $L$-input features $X_{L}$ satisfy
\begin{equation}
\sup_{\mathbf{u}}\|X_\infty(\mathbf{u})-X_L(\mathbf{u}_L)\|_2
 \leq e_X(L)=K_1r^L+K_2r^{2L} .
\label{eq:9}
\end{equation}
The nonnegative constants $K_{1}$ and $K_{2}$ depend on the fixed configuration and normalization, while $r$ is chosen as in Lemma~\ref{lem:2}. Both feature vectors use the same recent $L$ inputs (Fig.~\ref{fig:2}(d)). The bound follows by summing the omitted responses in Eq.~\eqref{eq:7}, with the terms of order $r^{L}$ arising from means and cross terms involving old and recent inputs. This establishes fading memory, meaning that changes in distant inputs have vanishing influence on the current output. For frozen weights $w$, define the $L^{2}(P)$ output discrepancy $d_{L} = \|w^{\top} X_{\infty} - w^{\top} X_{L}\|_{L^{2}(P)}$, which satisfies $d_{L} \leq \|w\|_{2} e_{X}(L)$. A usable memory horizon therefore depends on the task and readout as well as physical decay. The full proof and constants are given in Appendix~\ref{app:A4}.

For continuous operation, let $\mathcal{D}_{\infty}$ be the $L^{2}(P)$ closure of the union of all $\mathcal{D}_{L}$, and let $\mathcal{S}_{\infty}$ be the span of the full-history feature coordinates. Equation~\eqref{eq:9} gives $\mathcal{S}_{\infty} \subseteq \mathcal{D}_{\infty}$. The space therefore contains functions of arbitrarily old inputs, but every function in it still couples at most two distinct times.

\refstepcounter{cor}\label{cor:1}
\noindent\textbf{Corollary~\thecor. A task outside the span of candidate history functions.} Under independent uniform inputs on $[-1, 1]$, the target $Y^{(3)} = (\sqrt{3} u_{t})(\sqrt{3} u_{t-1})(\sqrt{3} u_{t-2})$ has unit variance and is orthogonal to $\mathcal{D}_{\infty}$. Its minimum mean squared error (MSE) is $1$ for every linear readout of the raw first and second moments in this model. Each candidate history function misses at least one independent, centered factor of this target. The proof is given in Appendix~\ref{app:A6}. Capturing this task requires an enlarged interaction family, such as that produced by the polynomial readouts used in the Gaussian QRC universality construction \cite{Nokkala2021GaussianReservoir}.

To examine the decay scales of these bounds, we use six persistent modes and the four parameter pairs $(\beta, \theta) = (3, 1.1), (1.02, 1.1), (3, 1.8)$, and $(1.02, 1.8)$. Every configuration is evaluated on the same $20$ coupling realizations, so that comparisons are paired. The normalized response energy $p_{k} = \|A^{k} b\|_{2}^{2} / \|b\|_{2}^{2}$ characterizes propagation, while $d_{L}$ measures the effect of discarded history on a fixed readout.

\begin{figure}[tbp]
\centering
\manuscriptfigure{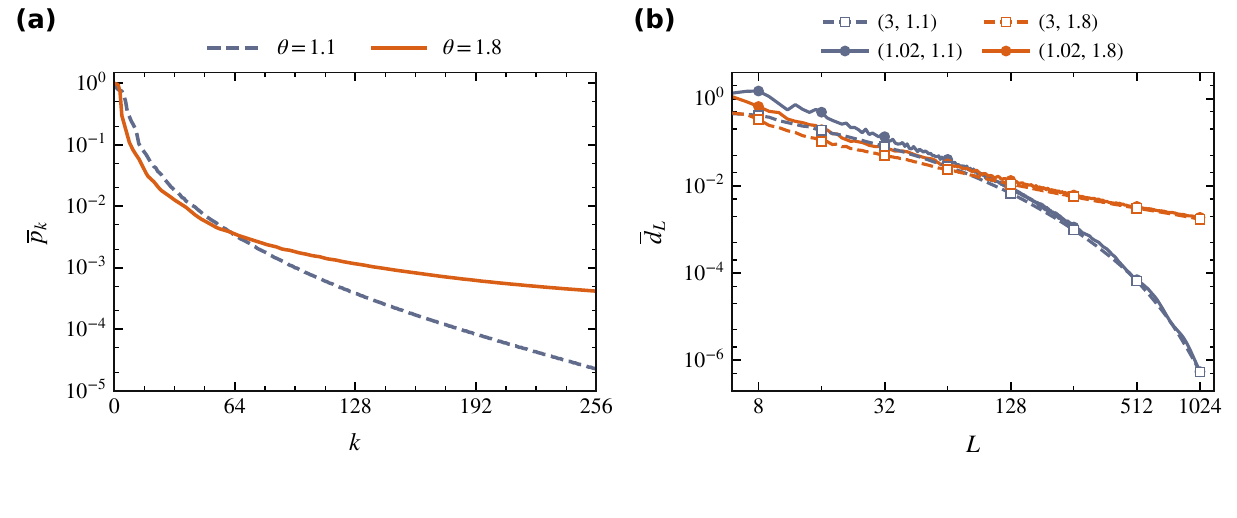}
\caption{\textbf{Memory response and finite-history approximation.} (a) Normalized response energy $p_{k} = \|A^{k} b\|_{2}^{2} / \|b\|_{2}^{2}$, where $k$ is input age, $A$ is the memory block of the single-step passive propagator, $b$ is its input-to-memory column, and $\|\cdot\|_{2}$ is the Euclidean norm. The propagator depends on $\theta$ and the static couplings. Its independence of the encoding parameter $\beta$ reduces the four configurations to two distinct physical responses. (b) Population output discrepancy $d_{L} = \|\hat{Y}_{\infty, t} - \hat{Y}_{L, t}\|_{L^{2}(P)}$, where $L$ is the number of retained recent inputs and $t$ denotes the observation step. The two outputs use the same frozen readout and the same recent inputs. The first starts from the stationary history-dependent memory state, while the second starts from vacuum at the beginning of the retained window. Readouts are the degree-three population teacher projections used in Fig.~\ref{fig:5}. The inputs are independent and uniform on $[-1, 1]$, and the $L^{2}(P)$ norm is evaluated under this distribution. The four $(\beta, \theta)$ configurations are distinguished in the legend. Overbars denote arithmetic averages over coupling seeds $0$ through $19$, with the same realizations as Figs.~\ref{fig:5} and~\ref{fig:6}. All sampled memory propagators have spectral radius below one. The quantity $d_{L}$ measures the approximation of a fixed predictor by a finite input history. The readout remains unchanged as $L$ varies.}
\label{fig:3}
\end{figure}

Figure~\ref{fig:3}(a) shows the propagation response independently of the forecasting target. The four encoding and propagation choices produce two distinct response families because $A$ and $b$ depend only on the propagation parameter $\theta$ and the static couplings. Changing $\beta$ modifies the input moment functions while leaving the normalized physical kernel unchanged at fixed $\theta$.

The mean response for $\theta = 1.8$ decreases more rapidly at short input ages, but its long tail decays more slowly than the response for $\theta = 1.1$. The maximum sampled spectral radii for $\theta = 1.1$ and $\theta = 1.8$ are approximately $0.990786$ and $0.999992$, respectively. All sampled propagators are stable. The near-unit spectral radius in part of the $\theta = 1.8$ ensemble is consistent with the observed slow tail. The numerical curves characterize the decay scales of these configurations, while the fading-memory result follows from the stability and continuity conditions established in Theorem~\ref{thm:2} and Appendix~\ref{app:A4}.

Figure~\ref{fig:3}(b) shows how this response translates into the output of a fixed predictor. We freeze the population readouts constructed for the controlled task in Sec.~\ref{sec:2.2} and retain the same recent inputs. The state at the start of the window is changed from the stationary memory state to vacuum. The resulting $d_{L}$ quantifies the change in output caused by this finite-history approximation. The ranking of the configurations changes with $L$, so the useful window depends on both the long-time response and the directions weighted by the readout.

The construction therefore provides an explicit list of the functions this reservoir can produce, namely single-input nonlinearities, delayed responses, and two-time products. Choosing a configuration then amounts to comparing this list with the history dependence of the target.

\section{Task alignment and prediction guarantees}\label{sec:2}

The readable history functions characterize the physical capabilities of the reservoir. We now describe the target on the same history variables and quantify its overlap with those functions. A \emph{teacher} $T$ approximates the predictable target while separating its nonlinear input dependence, delays, and temporal interactions. For a known task, $T$ is obtained analytically, whereas for forecasting data it is fitted on the training set. Holding this teacher fixed across candidate configurations ensures that performance differences reflect how well each readable space captures the same task.

\subsection{Describing task requirements with a teacher}

To identify which functions of the history the task requires, consider independent uniform inputs on $[-1, 1]$ and define $\psi_{n}(u) = \sqrt{2n + 1} P_{n}(u)$, where $P_{n}$ is the Legendre polynomial of degree $n$. The products $\Psi_{\boldsymbol{n}}(\mathbf{u}_{t, L}) = \prod_{k = 0}^{L - 1} \psi_{n_{k}}(u_{t - k})$ form an orthonormal basis for square-integrable functions of this history. Each nonzero entry of $\boldsymbol{n} = (n_{0}, \ldots, n_{L - 1})$ identifies a participating age and its polynomial degree, while the number of nonzero entries specifies the temporal interaction order.

\refstepcounter{lem}\label{lem:3}
\noindent\textbf{Lemma~\thelem. Teacher approximation.} Let $f_{L} = \mathbb{E}_{P}[Y \mid \mathbf{u}_{t, L}]$ and $\gamma_{\boldsymbol{n}} = \mathbb{E}_{P}[f_{L} \Psi_{\boldsymbol{n}}]$. For a finite index set $\mathcal{I}$, the teacher $T_{\mathcal{I}} = \sum_{\boldsymbol{n} \in \mathcal{I}} \gamma_{\boldsymbol{n}} \Psi_{\boldsymbol{n}}$ satisfies $\|f_{L} - T_{\mathcal{I}}\|_{2}^{2} = \sum_{\boldsymbol{n} \notin \mathcal{I}} \gamma_{\boldsymbol{n}}^{2}$. For each realization of estimated coefficients $\hat{\gamma}_{\boldsymbol{n}}$, define $\hat{T}_{\mathcal{I}} = \sum_{\boldsymbol{n} \in \mathcal{I}} \hat{\gamma}_{\boldsymbol{n}} \Psi_{\boldsymbol{n}}$. Orthonormality yields
\begin{equation}
\|f_L-\widehat T_{\mathcal I}\|_2^2
=\sum_{\boldsymbol n\notin\mathcal I}\gamma_{\boldsymbol n}^2
+\sum_{\boldsymbol n\in\mathcal I}
(\gamma_{\boldsymbol n}-\widehat\gamma_{\boldsymbol n})^2.
\end{equation}

The retained indices specify the task components, and their coefficients quantify the corresponding contributions under the independent input law (see Appendix~\ref{app:A7} for the proof).

\begin{table}[tbp]
\centering
\caption{\textbf{Task requirements and their physical counterparts.} Each requirement is described by the teacher and compared with the corresponding property of the reservoir.}
\label{tab:req}
\begingroup
\small
\setlength{\tabcolsep}{4pt}
\renewcommand{\arraystretch}{1.22}
\begin{tabularx}{\linewidth}{@{}>{\raggedright\arraybackslash}p{0.21\linewidth}
>{\raggedright\arraybackslash}p{0.29\linewidth} X@{}}
\toprule
\textbf{Task requirement} & \textbf{Teacher description} & \textbf{Physical comparison} \\
\midrule
Single-input nonlinearity & Polynomial degrees and coefficients at each age & Overlap with $\mu_{1}$ and
$\mu_{2}$ at the chosen $\beta$ \\
Delayed information & Participating ages and the residual from a finite history & Distinguishability of
the responses $v_{k}$, their conditioning, and the history tail \\
Cross-time interactions & Products at specified ages and their coefficients & Allowed products in
Eq.~\eqref{eq:7} and their readable combinations \\
\bottomrule
\end{tabularx}
\endgroup
\end{table}

The teacher thus maps each task requirement onto a direct comparison with the reservoir. For nonlinear input dependence, $\beta$ is assessed through overlaps with the fixed input moment functions in Eq.~\eqref{eq:4}, each of which combines several Fourier components. For delayed information, the allowable teacher approximation error sets the required history window, over which the responses $v_{k}$ must keep distinct delays distinguishable. For cross-time interactions, the raw moment model supplies products of first-moment components at two distinct ages. Under independent inputs, centered three-time products remain orthogonal to $\mathcal{D}_{\infty}$, as shown in Corollary~\ref{cor:1}.

For correlated time series, the Legendre products provide a common set of teacher features whose inner products are determined by the actual distribution $P$. We write $T = \boldsymbol{\gamma}^{\top} \Psi$ and retain the full Gram matrix of $\Psi$ in all projections, ensuring that the projection accounts for correlations across ages and polynomial terms.

\subsection{Candidate-function and alignment errors}\label{sec:2.2}

To identify which design change can reduce prediction error, we separate approximation within the candidate span $\mathcal{D}$ from realization within the readable space $\mathcal{S}$. Let $\mathcal{S} \subseteq \mathcal{D}$ denote either the finite-window or full-history construction, and let $f$ be the conditional mean given the corresponding history. For a finite window, $f = f_{L}$ and $\sigma_{\mathrm{irr}}^{2} = \sigma_{L}^{2} = \|Y - f_{L}\|_{2}^{2}$, which absorbs the uncertainty from omitted older inputs. Every projection is evaluated under the same distribution $P$.

\refstepcounter{thm}\label{thm:3}
\noindent\textbf{Theorem~\thethm. Task error decomposition.} For any predictor $g \in \mathcal{S}$, the risk $R(g) = \mathbb{E}_{P}[(Y - g)^{2}]$ decomposes as
\begin{equation}
\begin{aligned}
 R(g)&=\sigma_{\mathrm{irr}}^2
 +\underbrace{\|f-\Pi_{\mathcal{D}}f\|_2^2}_{\varepsilon_{\mathrm{cand}}^2}
 +\underbrace{\|\Pi_{\mathcal{D}}f-\Pi_{\mathcal{S}}f\|_2^2}_{\varepsilon_{\mathrm{align}}^2}
 +\underbrace{\|\Pi_{\mathcal{S}}f-g\|_2^2}_{\varepsilon_{\mathrm{read}}^2} .
\end{aligned}
\label{eq:10}
\end{equation}

The candidate-function error $\varepsilon_{\mathrm{cand}}$ measures the component of the target outside $\mathcal{D}$, while the alignment error $\varepsilon_{\mathrm{align}}$ captures the part of its candidate-space projection left inaccessible by propagation and measurement. The readout error $\varepsilon_{\mathrm{read}}$ measures how far the fitted predictor deviates from the optimal target projection in $\mathcal{S}$. Together with the conditional residual $Y - f$, these contributions are mutually orthogonal, yielding the exact equality in Eq.~\eqref{eq:10} (Appendix~\ref{app:A8}).

For fixed input moment functions and observation order, modifying the propagation, mode allocation, or observed coordinates adjusts $\mathcal{S}$ within the same $\mathcal{D}$. Task-aligned selection among these configurations can reduce $\varepsilon_{\mathrm{align}}$. In contrast, changing $\beta$ reshapes the input moment functions and alters both $\mathcal{D}$ and its physical realization. Capturing target components outside $\mathcal{D}$ requires modifying the encoding functions or permitted interactions, whereas readout fitting controls the final deviation within $\mathcal{S}$. The teacher consolidates these design trade-offs into a single computable objective.

\refstepcounter{lem}\label{lem:4}
\noindent\textbf{Lemma~\thelem. Task projection score.} For reservoir features $X$ and teacher features $\Psi$, define $G = \mathbb{E}_{P}[X X^{\top}]$, $H_{T} = \mathbb{E}_{P}[X \Psi^{\top}]$, and $M_{T} = \mathbb{E}_{P}[\Psi \Psi^{\top}]$. The minimum-norm weights for projecting $T = \boldsymbol{\gamma}^{\top} \Psi$ onto $\mathcal{S}$ are $w_{T} = G^{+} H_{T} \boldsymbol{\gamma}$, and the projection residual is
\begin{equation}
J_T=\|T-\Pi_{\mathcal{S}}T\|_2^2
 =\boldsymbol\gamma^\top
 (M_T-H_T^\top G^+H_T)\boldsymbol\gamma .
\end{equation}
Here, $+$ denotes the Moore--Penrose pseudoinverse. Nested projection yields
\begin{equation}
J_T
=\|T-\Pi_{\mathcal D}T\|_2^2
+\|\Pi_{\mathcal D}T-\Pi_{\mathcal S}T\|_2^2.
\end{equation}

Thus, $J_{T}$ combines the candidate-function and physical alignment losses for the teacher. For an orthonormal teacher basis, set $K_{T} = H_{T}^{\top} G^{+} H_{T}$. Its diagonal entries for centered basis functions correspond to their individual capacities, whereas $\boldsymbol{\gamma}^{\top} K_{T} \boldsymbol{\gamma}$ also incorporates overlaps between their projections. These cross terms and task coefficients explain why the full projection score carries information beyond an aggregate capacity sum. Appendix~\ref{app:A7} derives the Gram formula.

To resolve these losses for a target with known history dependence, we use independent uniform inputs on $[-1, 1]$ and combine a nonlinear delayed response with a two-time interaction,
\begin{equation}
Y_t=\sin(\pi u_{t-3})+\frac{3}{\sqrt2}u_{t-1}u_{t-5} .
\label{eq:12}
\end{equation}
The two components are orthogonal and each has variance $1/2$, so $Y_{t}$ has zero mean and unit variance. The teacher $T_{p}$ retains the sine expansion through Legendre degree $p$ along with the exact interaction term. Projecting $T_{p}$ onto the full-history readable space yields $\hat{Y}_{p} = \Pi_{\mathcal{S}_{\infty}} T_{p}$, and the same frozen degree-three readouts supply the finite-history comparison in Fig.~\ref{fig:3}(b). Stationary moment equations provide exact population expectations for all four parameter pairs, as detailed in Appendix~\ref{app:B1}.

Figure~\ref{fig:5}(a) applies Eq.~\eqref{eq:10} to the degree-three teacher projection, separating $\varepsilon_{\mathrm{cand}}^{2}$, $\varepsilon_{\mathrm{align}}^{2}$, and $\varepsilon_{\mathrm{read}}^{2}$. Changing the encoding substantially reduces the candidate-function contribution. At fixed $\beta$, this contribution is shared across all propagation choices because they operate within the same $\mathcal{D}_{\infty}$, leaving the component of the candidate-space target projection outside $\mathcal{S}_{\infty}$ as the principal remaining loss. In this decomposition, the encoding determines which input functions are generated, while propagation and observation govern how strongly each reaches the readout. Here, $\varepsilon_{\mathrm{read}}$ arises because the prediction is $\Pi_{\mathcal{S}_{\infty}} T_{3}$ whereas the optimal target prediction is $\Pi_{\mathcal{S}_{\infty}} Y$. Their squared separation is $\|\Pi_{\mathcal{S}_{\infty}}(Y - T_{3})\|_{2}^{2}$, matching the readout term in Eq.~\eqref{eq:10}. Using exact population moments and the full stationary history isolates the effect of teacher truncation on this readout.

\subsection{Prediction and selection guarantees}

To bound the prediction error on $Y$ via this proxy, let $\varepsilon_{\mathrm{app}} = \|f - T\|_{2}$ and $\delta_{T} = \|Y - T\|_{2}$. Since $T$ is a function of the available history, $\delta_{T}^{2} = \sigma_{\mathrm{irr}}^{2} + \varepsilon_{\mathrm{app}}^{2}$. The teacher projection $g_{T} = \Pi_{\mathcal{S}} T$ then satisfies
\begin{equation}
R(g_T)=R_*+\|\Pi_{\mathcal{S}}(f-T)\|_2^2,
 \qquad 0\leq R(g_T)-R_*\leq\varepsilon_{\mathrm{app}}^2 .
\label{eq:13}
\end{equation}
Equation~\eqref{eq:13} bounds the excess risk incurred by replacing the target with its teacher at a fixed configuration. It also separates teacher approximation from the structural restrictions of $\mathcal{S}$, since refining $T$ improves the task description while the optimal risk $R_{*}$ remains governed by the chosen readable space (Appendix~\ref{app:A8}).

For an empirically fitted predictor, two additional error sources arise, namely the readout's deviation from the teacher projection and the truncation of the memory response to a finite history. For a fitted full-history predictor $g_{\infty} = w^{\top} X_{\infty}$, define its deviation from the teacher projection as $\varepsilon_{\mathrm{fit}} = \|g_{\infty} - \Pi_{\mathcal{S}_{\infty}} T\|_{2}$. Freezing these weights and evaluating $g_{L} = w^{\top} X_{L}$ on the same $L$ recent inputs with vacuum memory at the window boundary gives, for $d_{L} = \|g_{\infty} - g_{L}\|_{2}$ and $J_{T}$ evaluated in $\mathcal{S}_{\infty}$,
\begin{equation}
\sqrt{R(g_L)}
 \leq\delta_T+\sqrt{J_T+\varepsilon_{\mathrm{fit}}^2}+d_L,
 \qquad d_L\leq\|w\|_2e_X(L) .
\label{eq:14}
\end{equation}
Here, $e_{X}(L)$ from Eq.~\eqref{eq:9} uniformly bounds the feature-vector truncation error over the retained $L$-input window. Orthogonality yields $\|T - g_{\infty}\|_{2}^{2} = J_{T} + \varepsilon_{\mathrm{fit}}^{2}$, and the triangle inequality then adds the teacher-to-target distance $\delta_{T}$ and the finite-history discrepancy $d_{L}$. Equation~\eqref{eq:14} thus combines three contributions, namely task approximation, realization error of the fitted predictor, and physical history truncation. Note that the teacher-relative deviation $\varepsilon_{\mathrm{fit}}$ and the target-relative error $\varepsilon_{\mathrm{read}}$ in Eq.~\eqref{eq:10} are measured against different reference projections.

\begin{figure}[tbp]
\centering
\manuscriptfigure{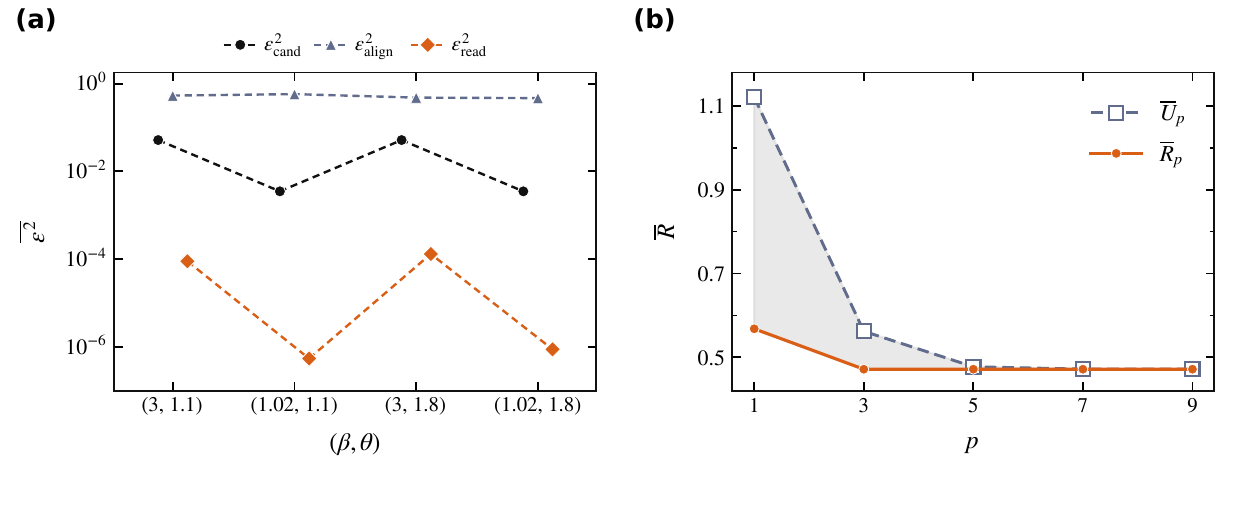}
\caption{\textbf{Population error decomposition and teacher bounds for a controlled history task.} Inputs $u_{t}$ are independent and uniformly distributed on $[-1, 1]$, and the target is $Y_{t} = \sin(\pi u_{t - 3}) + (3 / \sqrt{2}) u_{t - 1} u_{t - 5}$. The target has zero mean and unit variance. Each reservoir contains six persistent memory modes and uses the complete set of observables up to second order. (a) Squared candidate-function, alignment, and readout errors $\varepsilon_{\mathrm{cand}}^{2}$, $\varepsilon_{\mathrm{align}}^{2}$, and $\varepsilon_{\mathrm{read}}^{2}$ from Eq.~\eqref{eq:10} for the four parameter pairs $(\beta, \theta) = (3, 1.1), (1.02, 1.1), (3, 1.8)$, and $(1.02, 1.8)$. The prediction is the population least-squares projection of the degree-three teacher. The readout component is the squared distance from this prediction to the population projection of the true target. The three components sum to the risk of the teacher projection on the true target, up to the irreducible variance $\sigma_{\mathrm{irr}}^{2}$, which vanishes for this deterministic target. Horizontal offsets separate markers evaluated at the same configuration. (b) Teacher resolution for the fixed configuration $(\beta, \theta) = (1.02, 1.8)$. The teacher $T_{p}$ retains the Legendre expansion of the sine through degree $p$ and preserves the two-time product exactly. Its observable projection is $\hat{Y}_{p, t}$. The prediction risk is $R_{p} = \mathbb{E}[(Y_{t} - \hat{Y}_{p, t})^{2}]$, and the bound is $U_{p} = (\delta_{p} + \sqrt{J_{p}})^{2}$, where $\delta_{p}^{2} = \mathbb{E}[(Y_{t} - T_{p})^{2}]$ and $J_{p} = \mathbb{E}[(T_{p} - \hat{Y}_{p, t})^{2}]$. The expectation is over the specified input distribution. The inequality $R_{p} \leq U_{p}$ holds for each coupling realization. Overbars denote arithmetic averages of the individually evaluated quantities over $20$ paired coupling realizations with seeds $0$ through $19$. Stationary moment equations include the full input history and use exact observable expectations. The shaded separation between the two curves represents the difference between the averaged bound and risk.}
\label{fig:5}
\end{figure}

For the controlled task, define $R_{p} = \|Y - \hat{Y}_{p}\|_{2}^{2}$, $J_{p} = \|T_{p} - \hat{Y}_{p}\|_{2}^{2}$, and $\delta_{p} = \|Y - T_{p}\|_{2}$. Exact population projection and the full stationary history set $\varepsilon_{\mathrm{fit}} = d_{L} = 0$ in Eq.~\eqref{eq:14}, yielding $R_{p} \leq U_{p} = (\delta_{p} + \sqrt{J_{p}})^{2}$. Varying $p$ therefore tests how teacher resolution affects the prediction bound at a fixed physical configuration. Figure~\ref{fig:5}(b) examines this dependence at $(\beta, \theta) = (1.02, 1.8)$. Increasing the degree from $p = 3$ to $p = 7$ lowers the averaged bound from approximately $0.561943$ to $0.471577$, while the corresponding averaged population risks remain nearly unchanged at $0.471364$ and $0.471363$. This comparison shows that a more accurate teacher substantially tightens the bound around the attained risk, even though the risk itself varies little for this configuration. For each coupling realization, the bound is evaluated before averaging, preserving the per-realization inequality across the ensemble.

The same teacher-to-target distance $\delta_{T}$ also governs how accurately the projection score ranks candidate configurations. Consider configurations $\eta$ specifying the reference input state, encoding spectrum, propagation, persistent-mode allocation, and observed coordinates within a feasible set $\mathcal{A}$ that enforces the resource budget and stability requirements. With a common teacher and evaluation distribution, define $J_{\eta} = \|T - \Pi_{\mathcal{S}_{\eta}} T\|_{2}^{2}$. Minimizing this score prioritizes input functions and delayed combinations according to their contribution to the target.

\refstepcounter{cor}\label{cor:2}
\noindent\textbf{Corollary~\thecor. Accuracy of configuration selection.} Suppose estimates $\hat{d}_{T}(\eta)$ of $\sqrt{J_{\eta}}$ satisfy $\sup_{\eta \in \mathcal{A}} |\hat{d}_{T}(\eta) - \sqrt{J_{\eta}}| \leq \varepsilon_{T}$, where $\varepsilon_{T}$ is a uniform score error. If the selected configuration $\hat{\eta}$ minimizes the estimated score to tolerance $\gamma_{\mathrm{opt}} \geq 0$, then
\begin{equation}
\sqrt{R_*(\widehat\eta)}
 \leq\inf_{\eta\in\mathcal{A}}\sqrt{R_*(\eta)}
       +2\delta_T+2\varepsilon_T+\gamma_{\mathrm{opt}} .
\label{eq:15}
\end{equation}

Here, $R_{*}(\eta)$ denotes the minimum true-target risk in $\mathcal{S}_{\eta}$. For each fixed candidate $\eta$, $\sqrt{R_{*}(\eta)} = \operatorname{dist}(Y, \mathcal{S}_{\eta})$ and $\sqrt{J_{\eta}} = \operatorname{dist}(T, \mathcal{S}_{\eta})$. Because distance to a fixed subspace is $1$-Lipschitz in its argument, $|\sqrt{R_{*}(\eta)} - \sqrt{J_{\eta}}| \leq \delta_{T}$. Applying this inequality before and after score minimization yields Eq.~\eqref{eq:15}. This bound applies to ideal readable spaces under the stated uniform estimation condition, while Eq.~\eqref{eq:14} supplies the fitting and history terms for a realized predictor (Appendix~\ref{app:A9}).

\section{Configuration selection and Mackey--Glass forecasting}\label{sec:3}

We now apply the projection-score framework to empirical forecasting. After specifying the readout fitting and conditioning protocol, we isolate the effects of encoding and propagation on Mackey--Glass forecasting at a fixed resource budget and evaluate horizon-dependent selection alongside parameter transfer.

\subsection{Readout fitting and conditioning}

To compare selected configurations on equal terms, we apply the same feature preprocessing and fitting rule across all candidates. The fitted predictor is obtained from a training feature matrix $\mathsf{X}_{\mathrm{tr}}$, with one history per row, and a label vector $\boldsymbol{y}$. For $n_{\mathrm{tr}}$ training examples, we solve
\begin{equation}
\widehat w=\operatorname*{arg\,min}_w\left\{
 n_{\mathrm{tr}}^{-1}\|\boldsymbol{y}-\mathsf{X}_{\mathrm{tr}} w\|_2^2
 +\lambda_{\mathrm{rd}}w^\top D_0w\right\} .
\end{equation}
Here, $\lambda_{\mathrm{rd}} \geq 0$ is the ridge parameter, and $D_{0}$ is a positive semidefinite penalty matrix that leaves the intercept unpenalized. At $\lambda_{\mathrm{rd}} = 0$, we use the minimum-norm least-squares solution subject to any prescribed singular-value cutoff. Teacher outputs serve as fitting labels in the controlled proxy experiments and the Mackey--Glass forecasting experiments. For Mackey--Glass, the teacher is fitted using training data, and the reservoir readout is fitted to its outputs on the training histories; prediction errors are evaluated against the true future values. In the ETT and Weather experiments, the teacher guides memory construction, and the final readout is fitted to actual future values or increments according to the selected branch (Appendices~\ref{app:B3} and~\ref{app:B5}). Feature centering and scaling are fitted on the training set and then frozen. An invertible affine feature normalization preserves $\mathcal{S}$ when the intercept is included while altering the geometry of a regularized fit.

Weakly resolved feature directions can require large readout weights, increasing sensitivity to both feature perturbations and discarded history. For frozen weights $w$, a perturbation $\Delta X$ changes the prediction by at most $\|w\|_{2} \|\Delta X\|_{2}$. Consequently, regularization, singular-value truncation, and the bound $\|w\|_{2} e_{X}(L)$ enter the practical comparison alongside $J_{T}$. The fitting and numerical rank conventions are detailed in Appendix~\ref{app:B}.

\subsection{Encoding and propagation at fixed resources}

To test how the theoretical roles of encoding and propagation carry over to chaotic forecasting, we cross the STM-selected pair $(\beta, \theta) = (3, 1.1)$ with the single-step Mackey--Glass task-selected pair $(1.02, 1.8)$, yielding the intermediate configurations $(1.02, 1.1)$ and $(3, 1.8)$. All four combinations use six persistent modes, the same $90$ raw moment coordinates, the same readout fitting procedure, and $20$ paired coupling realizations. STM sums linear capacities over ages $1, \ldots, 20$ under independent uniform inputs, whereas forecasting error is evaluated on the Mackey--Glass series. Varying $\beta$ at fixed $\theta$ probes the input functions, while varying $\theta$ at fixed $\beta$ probes their propagation into readable history combinations.

The mean one-step normalized mean squared error (NMSE) drops from $1.9208 \times 10^{-5}$ for the STM-selected configuration to $2.8821 \times 10^{-7}$ for the jointly task-aligned configuration. Adjusting only the encoding reduces the mean test error by approximately $94.63$\%, whereas adjusting only the propagation at the original encoding reduces it by approximately $40.28$\%. Once the encoding is updated, modifying the propagation yields an additional reduction of approximately $72.05$\%. This dependence of the propagation gain on the encoding reflects the physical sequence established in Sec.~\ref{sec:history-functions}, where $\beta$ determines the accessible input functions and $\theta$ governs how effectively they reach the readout.

\begin{figure}[tbp]
\centering
\manuscriptfigure{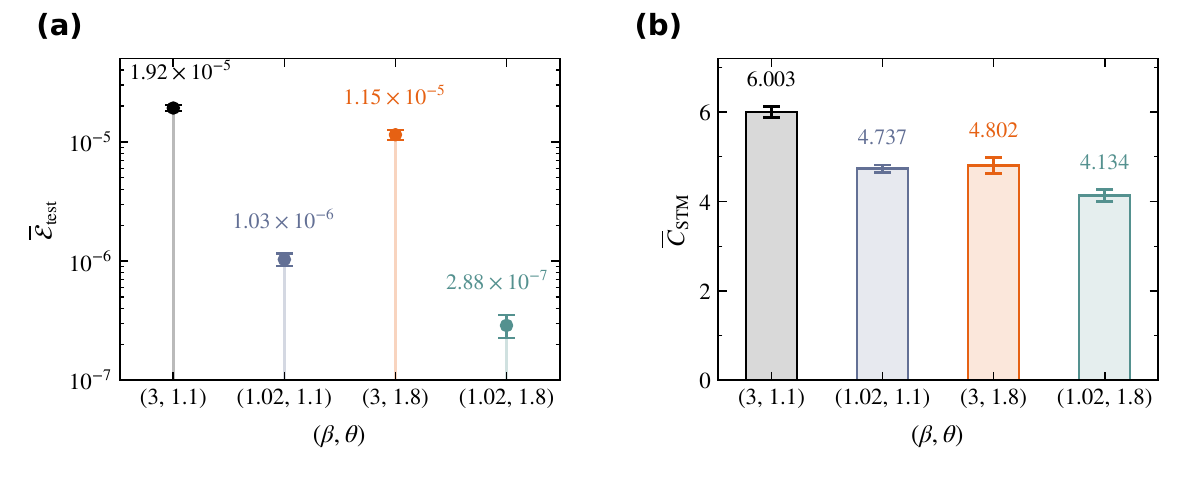}
\caption{\textbf{Task alignment at fixed quantum memory resources.} (a) Mean test normalized mean squared error for Mackey--Glass prediction one step ahead. (b) Mean linear short-term memory capacity for the same four configurations. The configuration pairs are $(\beta, \theta) = (3, 1.1), (1.02, 1.1), (3, 1.8)$, and $(1.02, 1.8)$, where $\beta$ controls the SNAP spectral shape and $\theta$ is the dimensionless propagation parameter. The first and last pairs are selected by STM maximization and task alignment, respectively. Crossing their encoding and propagation parameters isolates the effect of changing either parameter at fixed resources. All four configurations use six persistent memory modes, the same complete set of $90$ raw first- and second-moment coordinates, and the same prediction readout fitting procedure. STM capacities are evaluated analytically for independent inputs uniformly distributed on $[-1, 1]$ and summed over delays $1$ through $20$. Each mean uses the same $20$ paired static coupling realizations with seeds $0$ through $19$. Error bars in both panels show the mean plus or minus one standard error of the mean, computed as the sample standard deviation across these $20$ realizations divided by $\sqrt{20}$.}
\label{fig:6}
\end{figure}

Among these four configurations, the fully task-aligned choice achieves both the lowest forecasting error and the lowest summed linear STM capacity (Fig.~\ref{fig:6}). This result confirms that maximizing aggregate linear memory under independent uniform inputs does not guarantee strong performance on Mackey--Glass prediction, whereas task alignment selects configurations matched to the target's nonlinear and delayed structure.

\subsection{Horizon-dependent selection and parameter transfer}

\begin{figure}[tbp]
\centering
\includegraphics[width=\linewidth]{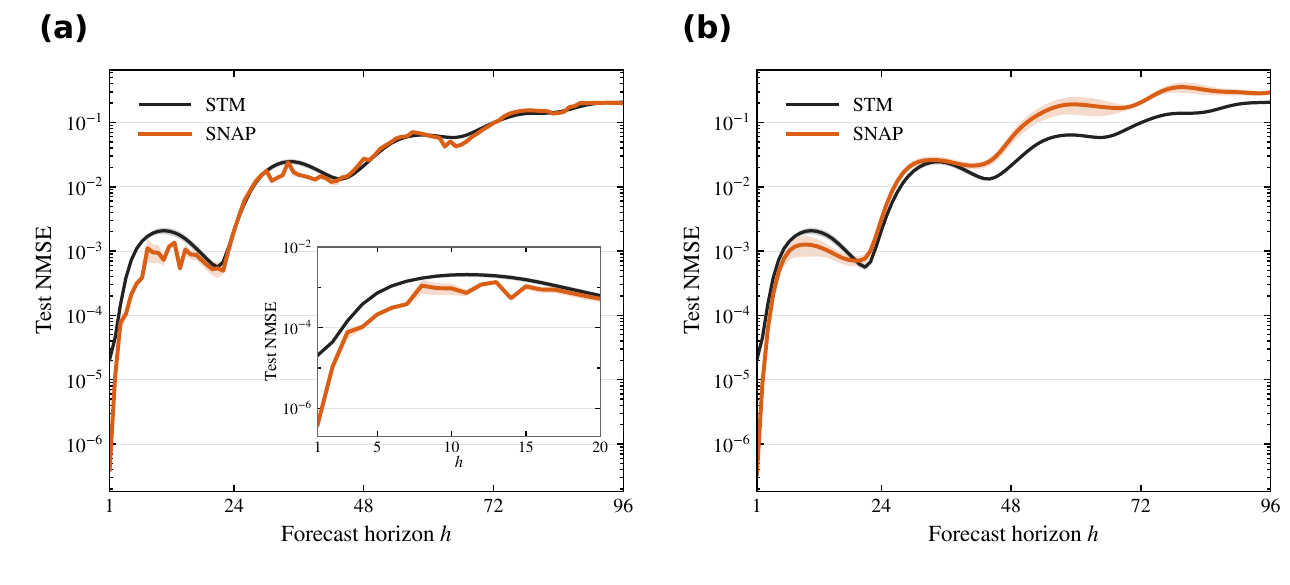}
\caption{\textbf{Horizon-dependent parameter selection and transfer.} Direct Mackey--Glass prediction errors for $h=1,\ldots,96$ data steps. (a) Task-aligned parameters $(\beta_h,\theta_h)$ are selected separately for each horizon using the training teacher projection score $J_h$; the inset enlarges $h=1,\ldots,20$. (b) The single-step pair $(\beta,\theta)=(1.02,1.8)$ is reused at every horizon. STM uses $(3,1.1)$ in both panels. Black and orange curves denote STM-based and task-aligned selection within the same SNAP reservoir family. Each evaluated configuration uses six persistent modes and $90$ raw moment coordinates. Both protocols fit a separate teacher target and readout for every horizon and share the teacher feature set, prediction origins, and paired coupling seeds. Test NMSE is normalized by the supervised training-target variance at each horizon. Curves show arithmetic means over $20$ coupling realizations; shading shows one standard error of the mean, computed as the sample standard deviation divided by $\sqrt{20}$. The two main panels share a logarithmic vertical scale, and the inset uses its own logarithmic scale. Protocol details are given in Appendix~\ref{app:B2}.}
\label{fig:7}
\end{figure}

The single-step comparison establishes how encoding and propagation affect forecasting at fixed resources. We next examine how the selected configuration performs as the forecast horizon changes. Figure~\ref{fig:7} compares two protocols for direct Mackey--Glass prediction over $h=1,\ldots,96$. In panel (a), the task-aligned pair $(\beta_h,\theta_h)$ is selected separately for each horizon using the training teacher projection score $J_h$. In panel (b), the single-step pair $(\beta,\theta)=(1.02,1.8)$ is reused across all horizons. The STM baseline uses $(3,1.1)$ in both panels. Every horizon has its own teacher target and fitted readout under both protocols. Each evaluated configuration uses six persistent modes, with common prediction origins and $20$ paired coupling realizations. Appendix~\ref{app:B2} details the teacher construction, parameter search, and fitting protocol.

With horizon-specific selection, task alignment achieves lower mean test NMSE than STM at every horizon from $1$ to $20$ (Fig.~\ref{fig:7}(a), inset). Averaging over these horizons reduces NMSE from $1.179\times10^{-3}$ to $6.302\times10^{-4}$, a decrease of $46.53\%$. At longer horizons, the curves converge and alternate in ranking. Across all $96$ horizons, the average NMSE is $0.06002$ for task alignment and $0.05970$ for STM, a difference of $0.54\%$. The forecasting benefit is therefore concentrated at short horizons, with comparable average performance over the complete range under the present teacher representation and parameter search.

Transferring the single-step configuration produces a different pattern (Fig.~\ref{fig:7}(b)). Its mean NMSE is below the STM baseline for $h=1,\ldots,19$ and above it for $h=20,\ldots,96$. This reversal occurs even though the readout is refitted for every horizon. It shows that a configuration selected for one-step prediction can still incur a forecasting penalty at longer horizons even after readout adaptation. Together, the two protocols support evaluating the physical configuration against the target at each forecast horizon, with the training teacher score and the resulting test performance assessed separately.

\section{Shared memory for multichannel forecasting}

Several outputs and forecast horizons can require overlapping history functions. A shared reservoir can supply these functions to independent readouts, so allocating persistent memory becomes part of the design problem.

\subsection{Task requirements under a shared memory budget}

Let $y_{t, c}$ be the observed value of output channel $c$, with forecast $\hat{y}_{t + h, c}$ at horizon $h$ from origin $t$. For $C_{\mathrm{out}}$ evaluated channels and maximum horizon $H_{\mathrm{pred}}$, the normalized mean squared errors are
\begin{equation}
\mathcal{E}_{c,h}=
 \frac{\sum_{t\in\mathcal{T}_{c,h}}
 (y_{t+h,c}-\widehat y_{t+h,c})^2}{N_{c,h}s_{c,h}^2},
 \qquad
 \mathcal{E}_{\mathrm{multi}}=\sum_{c=1}^{C_{\mathrm{out}}}
 \sum_{h=1}^{H_{\mathrm{pred}}}\omega_{c,h}\mathcal{E}_{c,h} .
\label{eq:17}
\end{equation}
Here, $\mathcal{T}_{c, h}$ is the set of scored forecast origins, $N_{c, h}$ is its size, and $s_{c, h}^{2}$ is the positive training reference variance. The nonnegative weights $\omega_{c, h}$ sum to one. For ETT and Weather, the reference is each channel's variance over the complete training segment, fixed across horizons. All outputs and all $96$ horizons receive equal weight, giving $\omega_{c, h} = 1 / (96 C_{\mathrm{out}})$. ETT has seven evaluated outputs and Weather has twenty.

For shared memory, let $T_{c, h}$ approximate the target for output $c$ and horizon $h$ divided by the positive training reference standard deviation $s_{c, h}$ used in Eq.~\eqref{eq:17}. Independent linear readouts use the same physical configuration, giving the selection objective
\begin{equation}
\begin{aligned}
J_\eta^{\mathrm{shared}}
&=\sum_{c,h}\omega_{c,h}
\|T_{c,h}-\Pi_{\mathcal S(\eta)}T_{c,h}\|_2^2,\\
\omega_{c,h}&\geq0,\qquad \sum_{c,h}\omega_{c,h}=1,\qquad
\sum_jm_j\leq M_{\mathrm{tot}}.
\end{aligned}
\end{equation}
Here, $m_{j}$ is the number of persistent modes in memory group $j$, and $M_{\mathrm{tot}}$ is the total budget. Each squared residual is measured relative to the training variance $s_{c, h}^{2}$, so equal weights $\omega_{c, h}$ assign equal importance to these normalized errors across outputs and horizons. The same projection and selection argument carries over to the weighted product space. This objective allocates shared memory according to the delayed functions required across outputs and horizons.

\subsection{Teacher-guided memory construction and selection}

To examine resource allocation across several inputs and outputs, we use ETT and Weather forecasting with shared passive memory. We call the six-mode model of Secs.~\ref{sec:1}--\ref{sec:3} the \emph{mechanism model}, and the shared-memory model of this section the \emph{application model}. A linear teacher fitted to future increments supplies task-relevant delay directions. Projecting the shift dynamics of the input delay embedding onto these directions defines a low-dimensional memory map. Candidate implementations use either a channel bank or four mixed input ports, with $8$ or $16$ persistent modes per port. Additional fresh vacuum ports complete the compressed map to a passive physical realization. The teacher determines the retained delay subspace, while aggregate validation error selects the physical candidate. Appendices~\ref{app:B3} and~\ref{app:B4} specify this construction.

Each persistent mode supplies the local observations $\langle \hat{q}_{j} \rangle$, $\langle \hat{p}_{j} \rangle$, $\langle \hat{q}_{j}^{2} \rangle$, and $\langle \hat{p}_{j}^{2} \rangle$, with $\hat{q}_{j} = (\hat{r}_{j} + \hat{r}_{j}^{\dagger}) / \sqrt{2}$ and $\hat{p}_{j} = (\hat{r}_{j} - \hat{r}_{j}^{\dagger}) / (i \sqrt{2})$. Appending the $C_{\mathrm{in}}$ standardized current inputs gives $4 m + C_{\mathrm{in}}$ features before the intercept. These quantum and classical coordinates jointly define the application's readable space, while the single-output reference implementations use the observation orders and additional memory branches listed in Appendix~\ref{app:B5}.

The final heads fit standardized future increments, which are added to the current value before restoring physical units. Ridge parameters and ranks of the forecast-horizon representation are selected per output, with chronological splits, fitting conventions, and persistent-mode counts specified in Appendix~\ref{app:B4}.

\subsection{Multichannel forecasting performance}

For shared multichannel prediction, the teacher supplies delay directions for passive memory, and validation selects among the resulting physical allocations. Each selected $M$ model uses one memory configuration across all evaluated outputs and $96$ horizons, with total persistent-mode counts listed in Table~\ref{tab:3} of Appendix~\ref{app:B}. Table~\ref{tab:1} reports these joint models alongside the single-output reference models $S$, which predict OT for ETT and air temperature for Weather.

For the four ETT multichannel comparisons, the reported mean NMSE of $M$ is below persistence. On ETTh2, its mean NMSE is $0.37826$, compared with $0.37997$ for Autoformer. On ETTm1 and ETTm2, its mean NMSE is lower than the reported Autoformer value by approximately $21.2\%$ and $23.1\%$, respectively. The reported multivariate linear and PatchTST mean errors are lower than those of $M$ on these four ETT datasets. On Weather, the $32$-mode shared model achieves a mean NMSE of $0.16674$, outperforming persistence as well as the PatchTST, DLinear, and Autoformer baselines.

\begin{figure}[tbp]
\centering
\manuscriptfigure{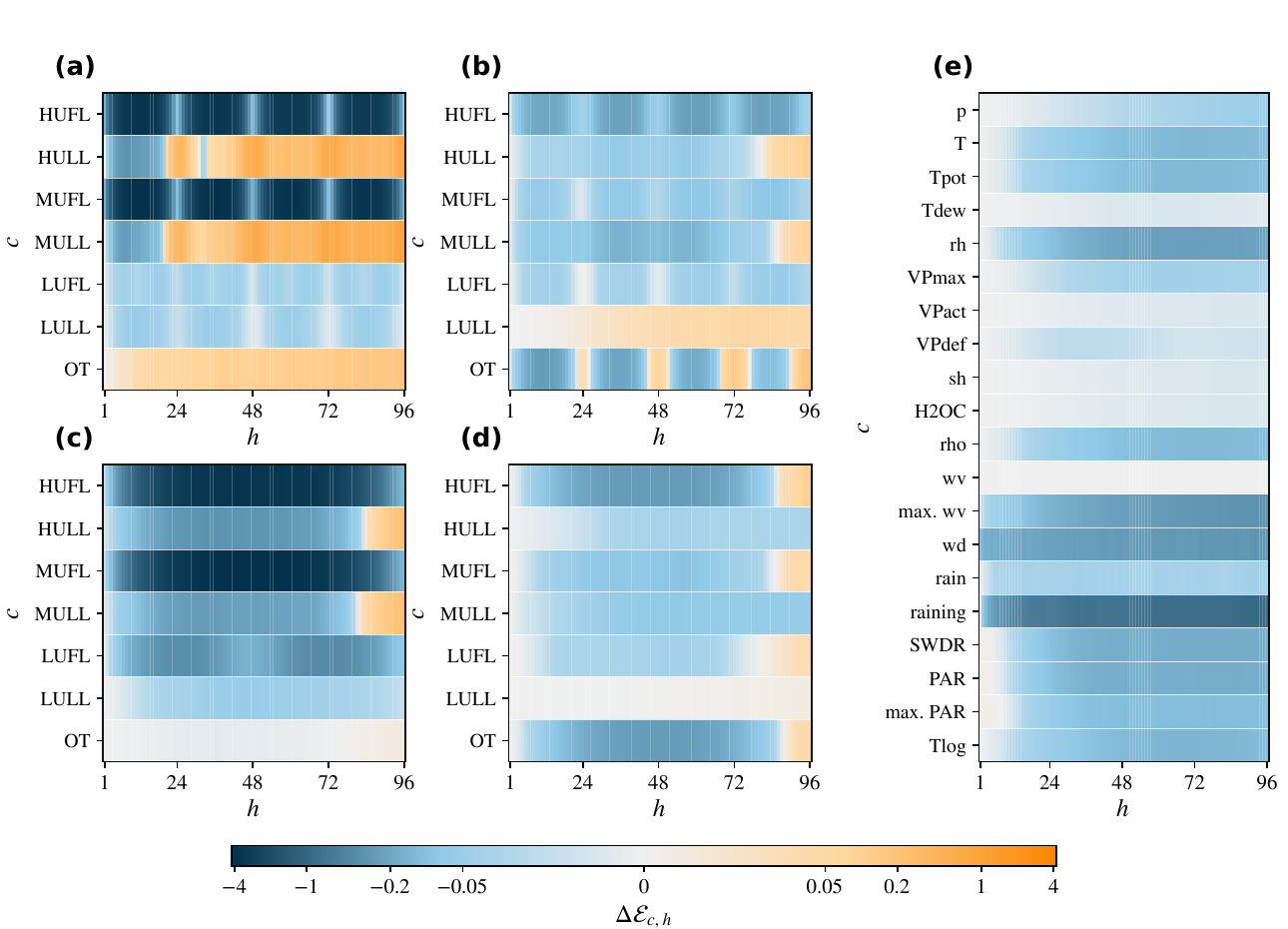}
\caption{\textbf{Output and horizon dependence of forecasting gains.} Panels (a) through (e) show ETTh1, ETTh2, ETTm1, ETTm2, and Weather, respectively. The horizontal coordinate $h = 1, \ldots, 96$ is the forecast horizon in data rows, and $c$ indexes the output channel. Each cell reports $\Delta \mathcal{E}_{c, h} = \mathcal{E}_{c, h}^{\mathrm{QRC}} - \mathcal{E}_{c, h}^{P}$, where $\mathcal{E}_{c, h}^{\mathrm{QRC}}$ and $\mathcal{E}_{c, h}^{P}$ are the test NMSEs of the selected joint SNAP QRC model and persistence. Persistence repeats the current observed value at every horizon. Errors are normalized by the variance of each output channel over the complete training segment. Negative values indicate lower error for QRC, positive values indicate higher error, and zero indicates equal errors. All panels share a symmetric logarithmic scale with a linear region for $|\Delta \mathcal{E}_{c, h}| \leq 0.02$. Each ETT panel includes all seven outputs, and Weather includes $20$ distinct outputs. The nominal sampling intervals are one hour for ETTh1 and ETTh2, $15$ minutes for ETTm1 and ETTm2, and $10$ minutes for Weather. The joint models are selected using validation data and evaluated on held-out test segments with seed $2021$.}
\label{fig:9}
\end{figure}

Figure~\ref{fig:9} breaks down the joint-model comparison with persistence by output and forecast horizon. The signed difference $\Delta \mathcal{E}_{c,h}$ reveals channel-dependent gains linked to the temporal structure of the data. The series exhibit pronounced daily oscillations in HUFL and MUFL for ETTh1 and ETTm1, with test correlations at a one-day lag of approximately $0.78$. Holding the current value fixed misses the within-day excursions of these channels, whereas a predictor using temporal features can exploit their recurring structure, providing a plausible explanation for the sustained blue regions. ETTh1 OT varies more slowly over short horizons, with a one-step test correlation of $0.983$ and a test variance equal to only $11.8\%$ of its training variance. Persistence is consequently a strong baseline in this test period, with a horizon-averaged NMSE of $0.069$, compared with $0.164$ for the fitted model. The orange regions in HULL and MULL also coincide with changes in their dynamics between data splits. For ETTh1, their test variances fall to approximately $35\%$ and $28\%$ of the corresponding training variances, while their one-day correlations decrease from $0.85$ and $0.87$ in training to $0.63$ and $0.61$ in testing. Both channels improve over persistence on the validation set but deteriorate on the test set. The reduced fluctuation amplitude strengthens persistence, while the altered daily dependence can limit the transferability of corrections learned from earlier histories, offering a data-based interpretation of their long-horizon losses. Several Weather channels retain or expand their gains at longer horizons, reflecting a widening error gap relative to persistence. Taken together, these post-hoc patterns suggest that the value of the learned temporal features depends on both the size of departures from the current value and the stability of those dependencies across data splits.

\begin{table}[p]
\centering
\caption{\textbf{Single-channel and multichannel forecasting performance.} (a) The single-channel setting uses SNAP QRC $(S)$, with OT as the target for the four ETT datasets and air temperature as the target for Weather. (b) The multichannel setting uses the archived joint SNAP QRC $(M)$ model, with seven input and evaluated output channels for ETT and $20$ for Weather. The symbols $C_{\mathrm{in}}$ and $C_{\mathrm{out}}$ denote the numbers of input and evaluated output channels. For $M$, $m$ counts the total persistent quantum memory modes, while fresh input and vacuum ports are accounted for separately. The $S$ references have a $32$-mode core with the branch-specific components and observation orders specified in Appendix~\ref{app:B5}. All reported errors are test NMSEs normalized by each output channel's variance over the complete training segment. The horizon $h$ is measured in data rows. Mean NMSE averages equally over $h = 1, \ldots, 96$ and, in the multichannel settings, over all output channels. Persistence repeats the current observed value at every horizon. Blank entries in the quantum mode column indicate that persistent quantum memory modes are not applicable to the corresponding classical baseline. Bold numerical entries identify the lowest available error within the same dataset, channel setting, and metric.}
\label{tab:1}
\begingroup
\fontsize{9.2}{11.2}\selectfont
\setlength{\tabcolsep}{4pt}
\renewcommand{\arraystretch}{1.13}
\begin{tabular*}{\linewidth}{@{\extracolsep{\fill}}lclrrr@{}}
\toprule
\multicolumn{6}{l}{\textbf{(a) Single-channel forecasting}} \\
\addlinespace[4pt]
\textbf{Dataset} & $C_{\mathrm{in}}\!\to\!C_{\mathrm{out}}$ & \textbf{Method} & $h=1$ & $h=96$ &
\textbf{Mean NMSE} \\
\midrule
\textbf{ETTh1} & $1 \to 1$ & Persistence & 0.00415 & 0.08900 & 0.06926 \\
 &  & \textbf{SNAP QRC (S)} & \textbf{0.00382} & 0.07664 & 0.05496 \\
 &  & PatchTST & 0.00435 & \textbf{0.07415} & \textbf{0.05399} \\
 &  & DLinear & 0.02202 & 0.09515 & 0.06608 \\
 &  & Autoformer & 0.07691 & 0.11316 & 0.09102 \\
\addlinespace[5pt]
\textbf{ETTh2} & $1 \to 1$ & Persistence & 0.01125 & 0.19943 & 0.29548 \\
 &  & \textbf{SNAP QRC (S)} & 0.01054 & 0.18178 & 0.13899 \\
 &  & PatchTST & \textbf{0.00858} & \textbf{0.17703} & \textbf{0.12971} \\
 &  & DLinear & 0.05147 & 0.20787 & 0.14778 \\
 &  & Autoformer & 0.07192 & 0.18798 & 0.15142 \\
\addlinespace[5pt]
\textbf{ETTm1} & $1 \to 1$ & Persistence & 0.00104 & 0.04619 & 0.03326 \\
 &  & \textbf{SNAP QRC (S)} & \textbf{0.00102} & 0.05129 & 0.02941 \\
 &  & PatchTST & 0.00107 & \textbf{0.04136} & \textbf{0.02586} \\
 &  & DLinear & 0.00198 & 0.04499 & 0.02785 \\
 &  & Autoformer & 0.03560 & 0.09841 & 0.07131 \\
\addlinespace[5pt]
\textbf{ETTm2} & $1 \to 1$ & Persistence & 0.00083 & 0.09447 & 0.22766 \\
 &  & \textbf{SNAP QRC (S)} & \textbf{0.00024} & \textbf{0.08493} & \textbf{0.06081} \\
 &  & PatchTST & 0.00071 & 0.08794 & 0.06490 \\
 &  & DLinear & 0.00208 & 0.08973 & 0.06534 \\
 &  & Autoformer & 0.07292 & 0.16288 & 0.13898 \\
\addlinespace[5pt]
\textbf{Weather} & $1 \to 1$ & Persistence & 0.00058 & 0.24610 & 0.16614 \\
 &  & \textbf{SNAP QRC (S)} & \textbf{0.00043} & 0.13725 & 0.08068 \\
 &  & PatchTST & 0.00134 & 0.13112 & 0.08309 \\
 &  & DLinear & 0.00213 & \textbf{0.12757} & \textbf{0.07790} \\
 &  & Autoformer & 0.17442 & 0.31616 & 0.22410 \\
\bottomrule
\end{tabular*}
\endgroup
\end{table}

\begin{table}[p]
\centering
\ContinuedFloat
\caption[]{\textbf{Single-channel and multichannel forecasting performance (continued).} (b) Multichannel forecasting. The mode count $m$ is the total persistent quantum memory allocated to the shared model. Blank entries in this column indicate that quantum mode counts are not applicable to the classical baselines.}
\begingroup
\fontsize{9.2}{11.2}\selectfont
\setlength{\tabcolsep}{4pt}
\renewcommand{\arraystretch}{1.13}
\begin{tabular*}{\linewidth}{@{\extracolsep{\fill}}lclrrrr@{}}
\toprule
\multicolumn{7}{l}{\textbf{(b) Multichannel forecasting}} \\
\addlinespace[4pt]
\textbf{Dataset} & $C_{\mathrm{in}}\!\to\!C_{\mathrm{out}}$ & \textbf{Method} & $m$ & $h=1$ & $h=96$ &
\textbf{Mean NMSE} \\
\midrule
\textbf{ETTh1} & $7 \to 7$ & Persistence &  & 0.17766 & 0.60362 & 1.29437 \\
 &  & Multivariate linear &  & 0.1183 & 0.5499 & 0.47277 \\
 &  & \textbf{SNAP QRC (M)} & 56 & 0.13596 & 0.82360 & 0.69294 \\
 &  & PatchTST &  & \textbf{0.11775} & \textbf{0.41346} & \textbf{0.37873} \\
 &  & DLinear &  & 0.22661 & 0.43472 & 0.39330 \\
 &  & Autoformer &  & 0.22592 & 0.55658 & 0.50243 \\
\addlinespace[5pt]
\textbf{ETTh2} & $7 \to 7$ & Persistence &  & 0.08510 & 0.46770 & 0.43166 \\
 &  & Multivariate linear &  & 0.0735 & 0.4448 & 0.33851 \\
 &  & \textbf{SNAP QRC (M)} & 32 & 0.08139 & 0.51568 & 0.37826 \\
 &  & PatchTST &  & \textbf{0.06681} & \textbf{0.35014} & \textbf{0.27467} \\
 &  & DLinear &  & 0.12467 & 0.43891 & 0.32336 \\
 &  & Autoformer &  & 0.24333 & 0.43506 & 0.37997 \\
\addlinespace[5pt]
\textbf{ETTm1} & $7 \to 7$ & Persistence &  & 0.05429 & 0.42103 & 1.21390 \\
 &  & Multivariate linear &  & 0.0467 & 0.3492 & 0.36987 \\
 &  & \textbf{SNAP QRC (M)} & 112 & 0.04839 & 0.42892 & 0.40631 \\
 &  & PatchTST &  & 0.04529 & \textbf{0.31903} & \textbf{0.29234} \\
 &  & DLinear &  & \textbf{0.04521} & 0.32532 & 0.30083 \\
 &  & Autoformer &  & 0.12915 & 0.59315 & 0.51577 \\
\addlinespace[5pt]
\textbf{ETTm2} & $7 \to 7$ & Persistence &  & 0.03485 & 0.26302 & 0.26646 \\
 &  & Multivariate linear &  & 0.0318 & 0.2368 & 0.17616 \\
 &  & \textbf{SNAP QRC (M)} & 32 & 0.03450 & 0.28218 & 0.20604 \\
 &  & PatchTST &  & \textbf{0.03066} & \textbf{0.22320} & \textbf{0.16352} \\
 &  & DLinear &  & 0.03333 & 0.23902 & 0.17103 \\
 &  & Autoformer &  & 0.20203 & 0.32711 & 0.26794 \\
\addlinespace[5pt]
\textbf{Weather} & $20 \to 20$ & Persistence &  & 0.03867 & 0.34754 & 0.25913 \\
 &  & \textbf{SNAP QRC (M)} & 32 & \textbf{0.03592} & \textbf{0.22525} & \textbf{0.16674} \\
 &  & PatchTST &  & 0.04328 & 0.22540 & 0.17058 \\
 &  & DLinear &  & 0.04265 & 0.23216 & 0.18089 \\
 &  & Autoformer &  & 0.32370 & 0.39469 & 0.37610 \\
\bottomrule
\end{tabular*}
\endgroup
\end{table}

\section{Conclusion and Discussion}

The results show that the predictive capability of a quantum reservoir depends on the functions accessible to its readout, rather than on aggregate computational capacity alone. This distinction is already apparent at the encoding stage. For the eight-level SNAP encoding considered here, the two nonlinear spectra with $\beta=1.02$ and $\beta=3$ generate Fourier spaces of the same dimension, while their first- and second-order input moments span four independent nonconstant functions with different shapes. Changing the encoding spectrum can therefore alter task-relevant representations without increasing the number of independent input functions. In the fixed-resource Mackey--Glass comparison, task-aligned encoding and propagation reduce the mean one-step prediction error by approximately a factor of $67$, despite yielding a lower summed linear STM capacity. The parameter controls further show that the effect of propagation depends on the input functions supplied by the encoding. In these experiments, alignment with the target was a better predictor of performance than total memory capacity.

The physical analysis explains how these task-specific differences arise. SNAP encoding determines the nonlinear functions introduced into the reservoir, while hopping dynamics distribute their contributions across memory modes and temporal delays. The selected observables then determine which combinations remain accessible to the readout. For raw first- and second-order moments, the resulting history functions contain nonlinear responses to individual inputs and interactions involving at most two distinct input times. This imposes a structural restriction that cannot be removed by increasing the number of memory modes alone. Conversely, the finite-window recovery result shows that sufficiently many controllable modes and complete moment observations can recover the available candidate functions. Memory resources determine how well the permitted functions can be retained and distinguished, whereas encoding and observation order determine which classes of functions are available. The fading-memory analysis adds a further constraint, since a slowly decaying physical response does not necessarily preserve the combinations most useful for prediction. The finite-history discrepancy in Fig.~\ref{fig:3} consequently depends on both propagation and the selected readout.

The prediction-error decomposition provides a direct interpretation of these physical restrictions. Candidate-function error measures the target components excluded by the available input functions and interaction structure, while alignment error measures the components that remain inaccessible after propagation and observation. The controlled experiment in Fig.~\ref{fig:5} separates these contributions under a common input distribution. Changing the encoding reduces the candidate-function error, whereas the remaining loss is dominated by the realization of task-relevant directions within the readable space. Increasing the resolution of the Legendre teacher substantially tightens the prediction bound but produces little change in the attained population risk at a fixed configuration. This separates the accuracy of the task description from the expressive limitations of the reservoir. The score therefore compares configurations by their relevance to a target, using the actual input distribution and readout geometry.

Task alignment also depends on the prediction horizon and the allocation of shared resources. The Mackey--Glass experiments show that configurations selected for individual horizons can improve short-horizon predictions, while transferring a configuration selected for one-step forecasting does not preserve its relative performance across longer horizons. The relevant historical dependencies therefore change with the prediction target. In multichannel forecasting, the selected shared-memory models improve the average error over persistence on the four ETT datasets, although their errors remain above those of the multivariate linear and PatchTST benchmarks, while on Weather the $32$-mode shared model achieves the lowest mean error among all evaluated methods. The channel- and horizon-resolved results further show that these improvements are not distributed uniformly across outputs. Unlike the single-output references, the shared models use teacher-constructed passive memory, local quadrature observations, and current-input features. Their performance shows that both memory directions and readable features matter under a shared budget, so extending the retained history alone is not enough.

These findings point to two directions for extending task-aligned reservoir design. The first is to enlarge the readable function space when the target contains interactions beyond those supported by the selected moment order, through higher-order observations or nonlinear readouts. The second is to incorporate the conditioning and measurement cost of the accessible directions into resource selection. The present moment analysis identifies which functions are accessible through ideal expectation values, whereas finite sampling determines how reliably these directions can be estimated. Combining task-dependent function selection with measurement-aware resource allocation would bring the present representation analysis closer to physical implementations.

In this study, we introduced a SNAP-encoded bosonic quantum reservoir and established how its encoding spectrum, input moments, memory propagation, and observable readout jointly determine the accessible history functions. This construction distinguishes the candidate functions generated by quantum encoding from those that can be realized through finite physical resources. By relating these functions to the predictable component of a target, we derived a task-dependent error decomposition and a projection criterion for reservoir configuration selection. Controlled experiments demonstrate that task-aligned configurations can substantially improve forecasting accuracy even when their aggregate memory capacity is lower. Forecasting across horizons and multiple output channels further shows that useful memory and representation resources depend on the target. The framework therefore offers a way to choose reservoir parameters from the functions a task requires, rather than from aggregate capacity alone.

\section*{Acknowledgments}
This work has been supported by the National Natural Science Foundation of China under Grant Nos. 62461160263 and 62371050. The funders had no role in study design, data collection and analysis, decision to publish, or preparation of the manuscript.

\bibliographystyle{unsrtnat}
\bibliography{references}

\appendix
\makeatletter
\renewcommand{\@seccntformat}[1]{%
  \expandafter\ifx\csname c@#1\endcsname\c@section
    Appendix~\thesection.\quad
  \else
    \csname the#1\endcsname\quad
  \fi
}
\makeatother

\numberwithin{equation}{section}
\renewcommand{\theequation}{\Alph{section}\arabic{equation}}
\numberwithin{table}{section}
\renewcommand{\thetable}{\Alph{section}\arabic{table}}

\section{Theoretical derivations and proofs}

We use a single fresh data mode and $m$ persistent canonical bosonic modes. The data mode is reset at every step. Conditional on the classical input history, each fresh mode is prepared independently of the retained quantum state. Classical independence between input times is imposed only where explicitly stated. The scalar input belongs to a compact interval $I$. Quantum expectation values are denoted by angle brackets, whereas $\mathbb{E}_{P}$ denotes expectation over a classical evaluation distribution $P$. For a real square-integrable function $f$, we denote its $L^{2}$ norm by $\|f\|_{2}$, with $\|f\|_{2}^{2} = \mathbb{E}_{P}[f^{2}]$. For deterministic vectors and matrices, $\|\cdot\|_{2}$ denotes the Euclidean norm and induced operator norm, respectively. The Frobenius norm is denoted by $\|\cdot\|_{\mathrm{F}}$. Subsections are ordered by proof dependency.

\subsection{Capacity sums and task direction}\label{app:A1}

Let $\{q_{a}\}_{a = 1}^{r_{\mathcal{S}}}$ be an orthonormal basis of the centered readable space. For centered orthonormal test functions $e_{j}$, projection gives $C(e_{j}) = \sum_{a} |\langle e_{j}, q_{a} \rangle_{P}|^{2}$. Bessel's inequality gives $\sum_{j} C(e_{j}) \leq r_{\mathcal{S}}$, with equality when the test family is complete for the readable space. The complete capacity sum measures the dimension of the readable space, whereas a specified target additionally assigns coefficients to its components. In Fig.~\ref{fig:1}, $\Pi_{\mathcal{S}_{A}} Y = 0.8 e_{1}$ and $\Pi_{\mathcal{S}_{B}} Y = 0.6 e_{2}$. Orthogonality gives squared residuals $0.36$ and $0.64$, respectively, while both capacity sums equal one.

\subsection{Encoding Fourier space and input moment functions}\label{app:A2}

This subsection proves Lemma~\ref{lem:1}.

The SNAP input is $|\psi(u)\rangle = \sum_{n = 0}^{d_{\mathrm{F}} - 1} c_{n} e^{i \lambda_{n} u} |n\rangle$ with $\sum_{n = 0}^{d_{\mathrm{F}} - 1} |c_{n}|^{2} = 1$. Here, $d_{\mathrm{F}}$ is the Fock support of the input, $c_{n}$ are fixed complex amplitudes, and $\lambda_{n}$ are real encoding eigenvalues. Its density matrix has entries $c_{n} c_{j}^{*} e^{i(\lambda_{n} - \lambda_{j}) u}$. Thus, every expectation value of an input-independent observable has frequency support contained in $\Omega_{\mathrm{state}} = \{\lambda_{n} - \lambda_{j} \mid c_{n} c_{j}^{*} \neq 0\}$, with $|\Omega_{\mathrm{state}}| \leq 1 + d_{\mathrm{F}}(d_{\mathrm{F}} - 1)$. If $I$ has nonempty interior, distinct exponential functions are linearly independent on $I$. When arbitrary Hermitian input observables are allowed, each nonzero frequency pair can be isolated by measuring the real and imaginary parts of $|n\rangle\langle j|$. The encoding Fourier space then has dimension $1 + 2 K_{\Omega} = |\Omega_{\mathrm{state}}|$, where $K_{\Omega}$ is the number of distinct positive frequencies.

The identity $\hat{a}_{0}^{p} |n + p\rangle = \sqrt{(n + p)! / n!}\, |n\rangle$ gives Eq.~\eqref{eq:4}. At fixed $\beta$, write $\mu(u) = \mu_{1}(u, \beta)$ and $\zeta(u) = \mu_{2}(u, \beta)$. These input moment functions sample adjacent-level gaps and gaps between levels separated by two, respectively. The number moment $\nu = \langle \hat{a}_{0}^{\dagger} \hat{a}_{0} \rangle_{u} = \sum_{n} n |c_{n}|^{2}$ is input independent.

Define $(\xi_{1}, \xi_{2}, \xi_{3}, \xi_{4}) = (\operatorname{Re}\mu, \operatorname{Im}\mu, \operatorname{Re}\zeta, \operatorname{Im}\zeta)$. Their nonconstant rank is $r_{\mathrm{src}}(\beta) = \dim\operatorname{span}_{\mathbb{R}} \{1, \xi_{1}, \xi_{2}, \xi_{3}, \xi_{4}\} - 1 \leq 4$. Each moment supplies a fixed weighted sum of its Fourier components. The encoding Fourier space measures how many frequency directions can be adjusted independently under arbitrary Hermitian input observations, while the input moment rank measures how many independent functions the selected moments supply.

For $d_{\mathrm{F}} = 8$ and $c_{n} = 1 / \sqrt{8}$, the normalized linear-spectrum limit has gaps that are integer multiples of $11 / 20$, giving seven distinct positive gaps and Fourier dimension $15$. At $\beta = 1.02$ and $\beta = 3$, enumeration of $\lambda_{j} - \lambda_{n}$ for $j > n$ gives $28$ distinct positive gaps and dimension $57$. The first and second moments contain seven and six positive frequencies, respectively, at these finite $\beta$ values. Their distinct frequency supports imply that $\{1, \operatorname{Re}\mu, \operatorname{Im}\mu, \operatorname{Re}\zeta, \operatorname{Im}\zeta\}$ is linearly independent on an input interval of nonzero length. In the linear limit, $\mu$ and $\zeta$ are nonzero multiples of $\exp(11 i u / 20)$ and $\exp(11 i u / 10)$, giving the same nonconstant input moment rank of four.

\subsection{Passive leakage, stability, and controllability}\label{app:A3}

This subsection proves Lemma~\ref{lem:2}.

For the Hermitian single-particle coupling matrix $\mathsf{h}$ and interaction time $\tau$, write the unitary mode propagation matrix as
\begin{equation}
e^{-i\tau\mathsf{h}}=
\begin{pmatrix}
s&\ell\\
b&A
\end{pmatrix}.
\end{equation}
The matrix $A$ acts on retained modes, $b$ is the data-to-memory column, and $\ell$ is the memory-to-reset-port row. The scalar $s$ acts within the reset port. Unitarity gives the block identities $A^{\dagger} A + \ell^{\dagger} \ell = I_{m}$ and $A A^{\dagger} + b b^{\dagger} = I_{m}$. For $v_{k} = A^{k} b$, direct substitution yields $\|v_{k + 1}\|_{2}^{2} = \|v_{k}\|_{2}^{2} - |\ell v_{k}|^{2}$. Hence, the response energy summed over all retained modes is nonincreasing with input age. The identity places no componentwise monotonicity requirement on $v_{k}$.

Since $A$ is a contraction, $\operatorname{spr}(A) \leq 1$. If $A x = \lambda x$ for a nonzero vector $x$ and $|\lambda| = 1$, then the first block identity gives $\ell x = 0$. It follows that $(0, x)^{\top}$ is an eigenvector of the full mode propagation matrix with eigenvalue $\lambda$. Conversely, an eigenvector of that matrix entirely supported on the retained modes gives a unit-modulus eigenvalue of $A$. Therefore, $\operatorname{spr}(A) < 1$ holds precisely when there is no such retained eigenmode.

For every $r$ satisfying $\operatorname{spr}(A) < r < 1$, there is a finite constant $C_{A}$ such that
\begin{equation}
\|A^k\|_2\leq C_A r^k
\qquad\text{for all }k\geq0.
\end{equation}
To see this, put $A$ in Jordan form. Each Jordan block contributes a polynomial in $k$ multiplied by a power of its eigenvalue. Because all eigenvalue moduli are strictly below $r$, these polynomial factors are bounded after division by $r^{k}$. The finitely many small values of $k$ can be included in $C_{A}$.

The second block identity also gives the telescoping relation
\begin{equation}
\sum_{k=0}^{L-1}v_kv_k^\dagger
=
I_m-A^L(A^\dagger)^L.
\end{equation}
Under stability, its limit is $I_{m}$. Thus, the vectors $\{A^{k} b\}_{k \geq 0}$ span the entire memory space. By the Cayley--Hamilton theorem, the first $m$ vectors already span this space.

\subsection{Moment recursions, fading memory, and history tails}\label{app:A4}

This subsection proves Theorem~\ref{thm:2}.

Let $\hat{\boldsymbol{r}}_{t}$ be the column of retained annihilation operators after the input at time $t$ has been injected and propagated. The passive evolution is $\hat{\boldsymbol{r}}_{t} = A \hat{\boldsymbol{r}}_{t - 1} + b \hat{a}_{0, t}$. Define $\alpha_{t} = \langle \hat{\boldsymbol{r}}_{t} \rangle$, $(Q_{t})_{ij} = \langle \hat{r}_{t, j}^{\dagger} \hat{r}_{t, i} \rangle$, and $(F_{t})_{ij} = \langle \hat{r}_{t, i} \hat{r}_{t, j} \rangle$. The matrix $Q_{t}$ is Hermitian and $F_{t}$ is complex symmetric. Write $\mu_{t} = \mu(u_{t})$ and $\zeta_{t} = \zeta(u_{t})$. Conditional product preparation implies that fresh-mode and memory expectations factor. Expanding the operator update therefore gives
\begin{equation}
\begin{aligned}
\alpha_t
&=A\alpha_{t-1}+b\mu_t,\\
Q_t
&=AQ_{t-1}A^\dagger+\nu bb^\dagger
  +A\alpha_{t-1}\mu_t^*b^\dagger
  +b\mu_t\alpha_{t-1}^\dagger A^\dagger,\\
F_t
&=AF_{t-1}A^\top+\zeta_tbb^\top
  +A\alpha_{t-1}\mu_tb^\top
  +b\mu_t\alpha_{t-1}^\top A^\top .
\end{aligned}
\end{equation}

Set $V_{t} = Q_{t} - \alpha_{t} \alpha_{t}^{\dagger}$ and $W_{t} = F_{t} - \alpha_{t} \alpha_{t}^{\top}$. The centered input functions are $\kappa(u) = \nu - |\mu(u)|^{2}$ and $\chi(u) = \zeta(u) - \mu(u)^{2}$. Subtraction of the mean outer products gives the closed affine recursions $V_{t} = A V_{t - 1} A^{\dagger} + \kappa(u_{t}) b b^{\dagger}$ and $W_{t} = A W_{t - 1} A^{\top} + \chi(u_{t}) b b^{\top}$. If two initial moment states receive the same inputs, their differences obey
\begin{equation}
\begin{aligned}
\Delta\alpha_t&=A^t\Delta\alpha_0,\\
\Delta V_t&=A^t\Delta V_0(A^\dagger)^t,\\
\Delta W_t&=A^t\Delta W_0(A^\top)^t.
\end{aligned}
\end{equation}
The power bound gives $O(r^{t})$ decay for the first moments and $O(r^{2t})$ decay for the centered second moments. Since the driven means are bounded, the identities $Q_{t} = V_{t} + \alpha_{t} \alpha_{t}^{\dagger}$ and $F_{t} = W_{t} + \alpha_{t} \alpha_{t}^{\top}$ give an $O(r^{t}) + O(r^{2t})$ bound for raw second-moment differences. The constants may depend on the finite initial moments.

For a history extending indefinitely into the past, the unique bounded moment response is
\begin{equation}
\begin{aligned}
\alpha_\infty
&=\sum_{k=0}^{\infty}v_k\mu(u_{t-k}),\\
V_\infty
&=\sum_{k=0}^{\infty}v_kv_k^\dagger\kappa(u_{t-k}),\\
W_\infty
&=\sum_{k=0}^{\infty}v_kv_k^\top\chi(u_{t-k}).
\end{aligned}
\label{eq:21}
\end{equation}

Let $M_{\mu} = \sup_{u \in I} |\mu(u)|$, $M_{\kappa} = \sup_{u \in I} |\kappa(u)|$, and $M_{\chi} = \sup_{u \in I} |\chi(u)|$. These quantities are finite because the input functions are continuous on $I$. With $c_{b} = C_{A} \|b\|_{2}$, define the response sums $H_{0} = \sum_{k = 0}^{\infty} \|v_{k}\|_{2}$, $H_{L} = \sum_{k = L}^{\infty} \|v_{k}\|_{2}$, and $T_{L} = \sum_{k = L}^{\infty} \|v_{k}\|_{2}^{2}$. They satisfy $H_{0} \leq c_{b} / (1 - r)$, $H_{L} \leq c_{b} r^{L} / (1 - r)$, and $T_{L} \leq c_{b}^{2} r^{2L} / (1 - r^{2})$. The finite-window moments $\alpha_{L}$, $V_{L}$, and $W_{L}$ are obtained by stopping the sums in Eq.~\eqref{eq:21} at $L - 1$. They are exactly the moments produced by the same $L$ inputs from vacuum memory, because the vacuum mean and its normally ordered centered moments vanish. Consequently,
\begin{equation}
\begin{aligned}
\|\alpha_\infty-\alpha_L\|_2
&\leq M_\mu H_L,\\
\|Q_\infty-Q_L\|_{\mathrm F}
&\leq M_\kappa T_L+2M_\mu^2H_0H_L,\\
\|F_\infty-F_L\|_{\mathrm F}
&\leq M_\chi T_L+2M_\mu^2H_0H_L.
\end{aligned}
\label{eq:22}
\end{equation}
For example, the second inequality follows from the tail of $V$ and $\|\alpha_{\infty} \alpha_{\infty}^{\dagger} - \alpha_{L} \alpha_{L}^{\dagger}\|_{\mathrm{F}} \leq (\|\alpha_{\infty}\|_{2} + \|\alpha_{L}\|_{2}) \|\alpha_{\infty} - \alpha_{L}\|_{2}$. Any fixed selection and linear normalization of raw moments defines a real feature vector $X$. Let $C_{\mathrm{obs}}$ bound this linear map with respect to the sum of the three moment norms in Eq.~\eqref{eq:22}. The constant feature, when included, has zero tail. An admissible uniform feature bound is
\begin{equation}
\begin{aligned}
\|X_\infty-X_L\|_2
&\leq C_{\mathrm{obs}}
\bigl[(M_\mu+4M_\mu^2H_0)H_L
 +(M_\kappa+M_\chi)T_L\bigr]\\
&\leq e_X(L)=K_1r^L+K_2r^{2L},
\end{aligned}
\label{eq:23}
\end{equation}
where $K_{1} = C_{\mathrm{obs}} (M_{\mu} + 4 M_{\mu}^{2} c_{b} / (1 - r)) c_{b} / (1 - r)$ and $K_{2} = C_{\mathrm{obs}} (M_{\kappa} + M_{\chi}) c_{b}^{2} / (1 - r^{2})$. These constants are fixed for a chosen physical configuration and feature scaling.

Every finite-window feature is continuous in finitely many inputs. The uniform tail bound therefore makes $X_{\infty}$ a uniform limit of continuous finite-history functions. More explicitly, equip bounded histories with $d_{\omega}(U, U^{\prime}) = \sup_{k \geq 0} \omega_{k} |u_{t - k} - u^{\prime}_{t - k}|$, where $\omega_{k} > 0$ and $\omega_{k} \to 0$. Choose $L$ so that both tails are small. For the remaining finitely many coordinates, a sufficiently small $d_{\omega}$ makes their feature difference small by continuity. This establishes fading memory of the moment features and of their linear readouts.

Finally, for a fixed real weight vector $w$,
\begin{equation}
|w^\top X_\infty-w^\top X_L|
\leq\|w\|_2e_X(L).
\label{eq:24}
\end{equation}
The same bound holds for the $L^{2}$ norm of the output difference.

\subsection{Candidate history functions, observable rank, and finite-window recovery}\label{app:A5}

This subsection proves Theorem~\ref{thm:1}.

For a vacuum-initialized window, write $\mu_{k} = \mu(u_{t - k})$ and $\zeta_{k} = \zeta(u_{t - k})$. Combining the centered series with the mean outer products gives
\begin{equation}
\begin{aligned}
\alpha_L
&=\sum_{k=0}^{L-1}v_k\mu_k,\\
Q_L
&=\nu\sum_{k=0}^{L-1}v_kv_k^\dagger
 +\sum_{\substack{0\leq k,l<L\\k\neq l}}
 v_kv_l^\dagger\mu_k\mu_l^*,\\
F_L
&=\sum_{k=0}^{L-1}v_kv_k^\top\zeta_k
 +\sum_{\substack{0\leq k,l<L\\k\neq l}}
 v_kv_l^\top\mu_k\mu_l .
\end{aligned}
\label{eq:25}
\end{equation}
The same-time products in the mean outer products cancel the corresponding terms in $\kappa$ and $\chi$. Thus, the candidate history functions of the raw-moment model include the constant, all $\xi_{j}(u_{t - k})$ for $j = 1, \ldots, 4$, and $\xi_{a}(u_{t - k}) \xi_{b}(u_{t - l})$ for $a, b \in \{1, 2\}$ and $k < l$.

Let $z_{L}$ collect the candidate history functions and define $\mathcal{D}_{L} = \operatorname{span}_{\mathbb{R}} z_{L}$ in $L^{2}(P)$. The number of listed coordinates is
\begin{equation}
N_L=1+4L+4\binom{L}{2}=1+2L^2+2L.
\end{equation}
Their independent dimension is $\dim\mathcal{D}_{L} = \operatorname{rank}\Gamma_{L} \leq N_{L}$, where $\Gamma_{L} = \mathbb{E}_{P}[z_{L} z_{L}^{\top}]$. Complete raw moments supply $n_{\mathrm{obs}} = 2m + m^{2} + m(m + 1) = 2m^{2} + 3m$ real coordinates from $\alpha$, $Q$, and $F$, respectively, together with a constant readout coordinate.

For a chosen observation set, Eq.~\eqref{eq:25} gives $X_{L} = \mathsf{B}_{L} z_{L}$, with $\mathsf{B}_{L}$ real. Its Gram matrix is $G_{L} = \mathbb{E}_{P}[X_{L} X_{L}^{\top}] = \mathsf{B}_{L} \Gamma_{L} \mathsf{B}_{L}^{\top}$. A coefficient vector lies in $\ker G_{L}$ exactly when its feature combination vanishes $P$-almost surely. Hence,
\begin{equation}
\dim\mathcal S_L
=\operatorname{rank}G_L
=\operatorname{rank}(\mathsf B_L\Gamma_L^{1/2})
\leq\min\{n_X,n_{\mathrm{obs}}+1,\dim\mathcal D_L\}.
\end{equation}
The Gram factor accounts for dependencies among functions under $P$, and $\mathsf{B}_{L}$ imposes the propagation and observation constraints.

For recovery, define the Krylov response matrix $\mathsf{K}_{L} = [b, Ab, \ldots, A^{L - 1} b]$. Suppose it has full column rank, and let $\mathsf{R}$ be a left inverse so that $\mathsf{R} \mathsf{K}_{L} = I_{L}$. At the operator level, $\hat{\boldsymbol{r}}_{t} = \mathsf{K}_{L} \hat{\boldsymbol{a}}_{t, L} + A^{L} \hat{\boldsymbol{r}}_{t - L}$, where $\hat{\boldsymbol{a}}_{t, L} = (\hat{a}_{0, t}, \ldots, \hat{a}_{0, t - L + 1})^{\top}$. For vacuum memory at $t - L$, all initial normally ordered moments and cross terms relevant here vanish. If $\boldsymbol{\mu}_{\mathrm{in}}$, $Q_{\mathrm{in}}$, and $F_{\mathrm{in}}$ denote the first and raw second moments of $\hat{\boldsymbol{a}}_{t, L}$, this gives $\mathsf{R} \alpha_{L} = \boldsymbol{\mu}_{\mathrm{in}}$, $\mathsf{R} Q_{L} \mathsf{R}^{\dagger} = Q_{\mathrm{in}}$, and $\mathsf{R} F_{L} \mathsf{R}^{\top} = F_{\mathrm{in}}$. Conditional product preparation gives diagonal entries $(Q_{\mathrm{in}})_{kk} = \nu$ and $(F_{\mathrm{in}})_{kk} = \zeta_{k}$, and off-diagonal entries $(Q_{\mathrm{in}})_{kl} = \mu_{k} \mu_{l}^{*}$ and $(F_{\mathrm{in}})_{kl} = \mu_{k} \mu_{l}$.

For completeness, write $\mu_{k} = x_{k} + i y_{k}$, and set $P_{kl} = \mu_{k} \mu_{l}^{*}$ and $T_{kl} = \mu_{k} \mu_{l}$. The real pair functions are recovered through
\begin{equation}
\begin{aligned}
x_kx_l&=\tfrac12(\operatorname{Re}P_{kl}+\operatorname{Re}T_{kl}),&
y_ky_l&=\tfrac12(\operatorname{Re}P_{kl}-\operatorname{Re}T_{kl}),\\
x_ky_l&=\tfrac12(\operatorname{Im}T_{kl}-\operatorname{Im}P_{kl}),&
y_kx_l&=\tfrac12(\operatorname{Im}T_{kl}+\operatorname{Im}P_{kl}).
\end{aligned}
\end{equation}
All these recovery operations are fixed linear combinations of measured raw moments. They recover every coordinate of $z_{L}$. Together with $\mathcal{S}_{L} \subseteq \mathcal{D}_{L}$, this proves $\mathcal{S}_{L} = \mathcal{D}_{L}$. For symmetrically ordered quadrature observations, the known commutator constants are removed through the intercept.

Under the single-port assumptions of Appendix~\ref{app:A3}, stability makes $\mathsf{K}_{m}$ invertible. Every prefix $\mathsf{K}_{L}$ is consequently full column rank for $L \leq m$. For the minimum-norm left inverse, $\|\mathsf{R}\|_{2} = 1 / \sigma_{\mathrm{min}}(\mathsf{K}_{L})$, where $\sigma_{\mathrm{min}}$ is the smallest singular value. Continuous operation retains the additional term $\mathsf{R} A^{L} \hat{\boldsymbol{r}}_{t - L}$. Moment reconstruction can amplify its first- and second-moment contributions by $\|\mathsf{R}\|_{2}$ and $\|\mathsf{R}\|_{2}^{2}$, respectively. Thus, finite-window exact recovery is an algebraic result whose continuous-operation error is governed by the tail and conditioning.

Define $\mathcal{D}_{\infty} = \operatorname{closure}_{L^{2}(P)}(\bigcup_{L \geq 1} \mathcal{D}_{L})$ and $\mathcal{S}_{\infty} = \operatorname{span}_{\mathbb{R}} \{X_{\infty, j}\}$. The uniform convergence in Eq.~\eqref{eq:23} implies $\mathcal{S}_{\infty} \subseteq \mathcal{D}_{\infty}$. This inclusion is the one needed for stationary calculations, in which features are evaluated at infinite history.

\subsection{Interaction order and nonuniversality}\label{app:A6}

This subsection proves Corollary~\ref{cor:1}.

Assume an independent uniform input process on $[-1, 1]$. Choose three distinct input ages $k_{1}$, $k_{2}$, and $k_{3}$, and let $Y_{t} = \prod_{j = 1}^{3} \sqrt{3}\, u_{t - k_{j}}$. Each factor has zero mean and unit variance. Independence gives $\mathbb{E}_{P}[Y_{t}] = 0$ and $\|Y_{t}\|_{2}^{2} = 1$.

Every candidate history function generating $\mathcal{D}_{L}$ depends on at most two input times. At least one factor in $Y_{t}$ is therefore independent of that function and has zero mean, which gives zero inner product. Continuity of the $L^{2}(P)$ inner product extends this orthogonality to $\mathcal{D}_{\infty}$ and hence to $\mathcal{S}_{\infty}$. Every allowed prediction $\hat{Y}$ satisfies $\|Y_{t} - \hat{Y}\|_{2}^{2} = 1 + \|\hat{Y}\|_{2}^{2} \geq 1$. The zero prediction attains equality.

The orthogonality persists as the encoding support, spectrum, number of modes, and passive couplings vary within the raw first- and second-moment architecture. The target is a continuous finite-memory functional and therefore has fading memory. Because its MSE is bounded below by one, a larger interaction family is needed to approximate general fading-memory targets.

\subsection{Legendre approximation and Gram projections}\label{app:A7}

This subsection proves Lemmas~\ref{lem:3} and~\ref{lem:4}.

For independent uniform inputs on $[-1, 1]$, define $\Psi_{\boldsymbol{n}}(U_{L}) = \prod_{k = 0}^{L - 1} \sqrt{2n_{k} + 1}\, P_{n_{k}}(u_{t - k})$ on $U_{L} = (u_{t}, \ldots, u_{t - L + 1})$, where $P_{n}$ is the degree-$n$ Legendre polynomial and $\boldsymbol{n} = (n_{0}, \ldots, n_{L - 1})$ has nonnegative integer entries. Univariate Legendre orthogonality and independence give orthonormality of this tensor basis. Polynomial completeness on the compact input cube gives completeness in $L^{2}(P)$.

For a square-integrable target $Y$, let $f_{L} = \mathbb{E}_{P}[Y \mid U_{L}]$. Its coefficients are $\gamma_{\boldsymbol{n}} = \mathbb{E}_{P}[f_{L} \Psi_{\boldsymbol{n}}]$. For a finite index set $\mathcal{I}$, the projection is $T_{\mathcal{I}} = \sum_{\boldsymbol{n} \in \mathcal{I}} \gamma_{\boldsymbol{n}} \Psi_{\boldsymbol{n}}$. If a fitted teacher has coefficients $\hat{\gamma}_{\boldsymbol{n}}$ on the same set, Parseval's identity and orthogonality give
\begin{equation}
\|f_L-\widehat T_{\mathcal{I}}\|_2^2
=
\sum_{\boldsymbol{n}\notin\mathcal{I}}\gamma_{\boldsymbol{n}}^2
+
\sum_{\boldsymbol{n}\in\mathcal{I}}
(\gamma_{\boldsymbol{n}}-\widehat \gamma_{\boldsymbol{n}})^2.
\label{eq:26}
\end{equation}

For fitted coefficients, the evaluation identities are conditional on the training data. The conditional-mean identity gives $\|Y - \hat{T}_{\mathcal{I}}\|_{2}^{2} = \sigma_{L}^{2} + \|f_{L} - \hat{T}_{\mathcal{I}}\|_{2}^{2}$, where $\sigma_{L}^{2} = \|Y - f_{L}\|_{2}^{2}$. For the controlled target $Y_{t} = \sin(\pi u_{t - 3}) + (3 / \sqrt{2}) u_{t - 1} u_{t - 5}$, the two centered components are orthogonal and each has variance $1 / 2$. The normalized sine coefficients are
\begin{equation}
\gamma_n=\frac{\sqrt{2n+1}}{2}
\int_{-1}^{1}\sin(\pi u)P_n(u)\,du.
\end{equation}
The teacher $T_{p}$ retains these coefficients through degree $p$ and the two-time product exactly. Consequently, $\delta_{p}^{2} = 1 / 2 - \sum_{n = 0}^{p} \gamma_{n}^{2}$, with $\gamma_{n} = 0$ for even $n$. This residual depends only on the target and input law and is common to every physical realization.

For a general input distribution, collect the chosen basis functions in $\Psi$, and set $M_{T} = \mathbb{E}_{P}[\Psi \Psi^{\top}]$. For any real target $g \in L^{2}(P)$, set $h_{T} = \mathbb{E}_{P}[\Psi g]$ and $\boldsymbol{\gamma}_{*} = M_{T}^{+} h_{T}$, where $+$ denotes the Moore--Penrose pseudoinverse. Every null vector of $M_{T}$ defines a zero function and is orthogonal to $h_{T}$, so $h_{T}$ belongs to the range of $M_{T}$. Completing the quadratic form gives
\begin{equation}
\|g-\boldsymbol\gamma^\top\Psi\|_2^2
=
\|g\|_2^2-h_T^\top M_T^+h_T
+(\boldsymbol\gamma-\boldsymbol\gamma_*)^\top M_T(\boldsymbol\gamma-\boldsymbol\gamma_*).
\end{equation}
This is the distribution-dependent replacement for a sum of squared coefficients.

Let $X$ be a real observed feature vector including the bias. For a teacher $T = \boldsymbol{\gamma}^{\top} \Psi$, define $G = \mathbb{E}_{P}[X X^{\top}]$ and $H_{T} = \mathbb{E}_{P}[X \Psi^{\top}]$. The population objective is $\|T - w^{\top} X\|_{2}^{2} = \boldsymbol{\gamma}^{\top} M_{T} \boldsymbol{\gamma} - 2 w^{\top} H_{T} \boldsymbol{\gamma} + w^{\top} G w$. The same null-space argument gives $H_{T} \boldsymbol{\gamma} \in \operatorname{range} G$. A minimum-norm optimizer is $w_{*} = G^{+} H_{T} \boldsymbol{\gamma}$, and the minimum is
\begin{equation}
J_T
=
\boldsymbol\gamma^\top(M_T-H_T^\top G^+H_T)\boldsymbol\gamma.
\end{equation}
For an orthonormal teacher basis, $M_{T} = I$ and $K_{T} = H_{T}^{\top} G^{+} H_{T}$ satisfies $0 \leq K_{T} \leq I$. Its diagonal entries for centered basis functions are their capacities. Its off-diagonal entries are inner products between the corresponding projected functions and contribute to $\boldsymbol{\gamma}^{\top} K_{T} \boldsymbol{\gamma}$. Thus, a general task mixture depends on both the capacities and the overlaps of its projected components. With $\mathcal{S} \subseteq \mathcal{D}$, Pythagoras also gives $J_{T} = \|T - \Pi_{\mathcal{D}} T\|_{2}^{2} + \|\Pi_{\mathcal{D}} T - \Pi_{\mathcal{S}} T\|_{2}^{2}$.

\subsection{Nested risk decompositions, teachers, and finite histories}\label{app:A8}

This subsection proves Theorem~\ref{thm:3} and the teacher bounds in Eqs.~\eqref{eq:13} and~\eqref{eq:14}.

Let $U$ denote the information available to the predictor, and let $f = \mathbb{E}_{P}[Y \mid U]$ for a square-integrable response $Y$. Write $\sigma_{\mathrm{irr}}^{2} = \|Y - f\|_{2}^{2}$. Let $\mathcal{S} \subseteq \mathcal{D}$ be closed real subspaces of square-integrable functions of $U$, and let $\Pi_{\mathcal{S}}$ and $\Pi_{\mathcal{D}}$ be their orthogonal projections. For any $\hat{Y} \in \mathcal{S}$, we write $f - \hat{Y} = (f - \Pi_{\mathcal{D}} f) + (\Pi_{\mathcal{D}} f - \Pi_{\mathcal{S}} f) + (\Pi_{\mathcal{S}} f - \hat{Y})$. The first term is orthogonal to $\mathcal{D}$. The second lies in $\mathcal{D}$ and is orthogonal to $\mathcal{S}$, since $\Pi_{\mathcal{S}} \Pi_{\mathcal{D}} f = \Pi_{\mathcal{S}} f$. The third lies in $\mathcal{S}$. The conditional-mean residual $Y - f$ is orthogonal to every function of $U$. Pythagoras therefore gives
\begin{equation}
\|Y-\widehat Y\|_2^2
=
\sigma_{\mathrm{irr}}^2
+\|f-\Pi_{\mathcal{D}}f\|_2^2
+\|\Pi_{\mathcal{D}}f-\Pi_{\mathcal{S}}f\|_2^2
+\|\Pi_{\mathcal{S}}f-\widehat Y\|_2^2.
\end{equation}
The spaces may be $(\mathcal{D}_{L}, \mathcal{S}_{L})$ for a finite window or $(\mathcal{D}_{\infty}, \mathcal{S}_{\infty})$ for full-history features, with $U$ chosen accordingly. For a deterministic task measurable with respect to $U$, $\sigma_{\mathrm{irr}} = 0$.

For finite-history features, apply the same decomposition to $f_{L}$ and use the conditional-mean identity. Writing
\begin{equation}
\begin{aligned}
\varepsilon_{\mathrm{cand}}
&=\|f_L-\Pi_{\mathcal{D}_L}f_L\|_2,\\
\varepsilon_{\mathrm{align}}
&=\|\Pi_{\mathcal{D}_L}f_L-\Pi_{\mathcal{S}_L}f_L\|_2,\\
\varepsilon_{\mathrm{read}}
&=\|\Pi_{\mathcal{S}_L}f_L-\widehat Y_L\|_2,
\end{aligned}
\end{equation}
one obtains $R_{L} = \sigma_{L}^{2} + \varepsilon_{\mathrm{cand}}^{2} + \varepsilon_{\mathrm{align}}^{2} + \varepsilon_{\mathrm{read}}^{2}$, where $R_{L} = \|Y - \hat{Y}_{L}\|_{2}^{2}$. The readout term describes the actual predictor's deviation from the optimal target projection, and its origin depends on the training protocol.

For a teacher $T$ that is a function of the available history, let $\hat{Y}_{T} = \Pi_{\mathcal{S}} T$, $\delta_{T} = \|Y - T\|_{2}$, $J_{T} = \|T - \Pi_{\mathcal{S}} T\|_{2}^{2}$, and $R_{*} = \|Y - \Pi_{\mathcal{S}} f\|_{2}^{2}$. Then
\begin{equation}
\|Y-\widehat Y_T\|_2^2
=R_*+\|\Pi_{\mathcal S}(f-T)\|_2^2,
\qquad
0\leq\|Y-\widehat Y_T\|_2^2-R_*
\leq\|f-T\|_2^2.
\end{equation}
The optimal residual $Y - \Pi_{\mathcal{S}} f$ is orthogonal to $\mathcal{S}$, while $\Pi_{\mathcal{S}}(f - T)$ lies in $\mathcal{S}$. Pythagoras gives the equality, and contractivity of projection gives the upper bound in Eq.~\eqref{eq:13}. The triangle inequality also yields
\begin{equation}
\|Y-\widehat Y_T\|_2
\leq\delta_T+\sqrt{J_T}.
\end{equation}
A truncated teacher can therefore produce a readout deviation through $\Pi_{\mathcal{S}}(f - T)$, even in an exact population calculation.

For the general teacher bound, let $T$ be any square-integrable teacher and set $\delta_{T} = \|Y - T\|_{2}$. For full-history features, write $\mathcal{S} = \mathcal{S}_{\infty}$, $g = \hat{w}^{\top} X_{\infty}$, $J_{T} = \|T - \Pi_{\mathcal{S}} T\|_{2}^{2}$, and $\varepsilon_{\mathrm{fit}} = \|g - \Pi_{\mathcal{S}} T\|_{2}$. The coefficient vector $\hat{w}$ is fixed during evaluation. Orthogonality within the teacher projection gives $\|T - g\|_{2}^{2} = J_{T} + \varepsilon_{\mathrm{fit}}^{2}$. Evaluate the same weights on the vacuum-window features to obtain $g_{L} = \hat{w}^{\top} X_{L}$, and define $R_{\infty} = \|Y - g\|_{2}^{2}$, $R_{L} = \|Y - g_{L}\|_{2}^{2}$, and $d_{L} = \|g - g_{L}\|_{2}$. These outputs are evaluated on coupled histories with the same retained recent inputs. The reverse triangle inequality and Eq.~\eqref{eq:24} give $|\sqrt{R_{\infty}} - \sqrt{R_{L}}| \leq d_{L} \leq \|\hat{w}\|_{2} e_{X}(L)$. The triangle inequality applied to $Y - g_{L} = (Y - T) + (T - g) + (g - g_{L})$ proves
\begin{equation}
\sqrt{R_L}
\leq
\delta_T+\sqrt{J_T+\varepsilon_{\mathrm{fit}}^2}+d_L
\leq
\delta_T+\sqrt{J_T+\varepsilon_{\mathrm{fit}}^2}
+\|\widehat w\|_2e_X(L).
\end{equation}
This definition of $\delta_{T}$ already includes response noise. For a teacher based on $U_{L}$, the conditional-mean identity gives $\delta_{T}^{2} = \sigma_{L}^{2} + \|f_{L} - T\|_{2}^{2}$, and Eq.~\eqref{eq:26} supplies the second term under its orthonormal-input assumptions.

For a stationary population teacher predictor, use $J_{p} = \|T_{p} - \Pi_{\mathcal{S}_{\infty}} T_{p}\|_{2}^{2}$ and $\delta_{p} = \|Y - T_{p}\|_{2}$. Freezing that predictor's weights and truncating its history gives the additional bound $R_{L, p} \leq (\delta_{p} + \sqrt{J_{p}} + d_{L, p})^{2}$, where $d_{L, p}$ is the discrepancy between the stationary and vacuum-window outputs with those same weights.

All these bounds are pointwise in the physical configuration. If $\phi = 1, \ldots, N$ indexes the $N$ coupling realizations, define $U_{p, \phi} = (\delta_{p} + \sqrt{J_{p, \phi}})^{2}$ before averaging. Summing $R_{p, \phi} \leq U_{p, \phi}$ gives $\bar{R}_{p} \leq \bar{U}_{p}$, where bars denote arithmetic means over $\phi$. For a common teacher, $\delta_{p}$ is independent of $\phi$, and $\bar{U}_{p} = \delta_{p}^{2} + \bar{J}_{p} + 2 \delta_{p} \operatorname{avg}(\sqrt{J_{p}}) \leq (\delta_{p} + \sqrt{\bar{J}_{p}})^{2}$. The last inequality follows from concavity of the square root.

\subsection{Configuration selection and shared output objectives}\label{app:A9}

This subsection proves Corollary~\ref{cor:2}.

Let $\mathcal{A}$ be the candidate set after imposing the physical resource budget. For $\eta \in \mathcal{A}$, let $R_{\eta}$ be the population evaluation risk and $\hat{R}_{\eta}$ its estimate. Suppose $\sup_{\eta \in \mathcal{A}} |\hat{R}_{\eta} - R_{\eta}| \leq \varepsilon$. If an approximate selector satisfies $\hat{R}_{\hat{\eta}} \leq \inf_{\eta \in \mathcal{A}} \hat{R}_{\eta} + \gamma_{\mathrm{opt}}$, where $\gamma_{\mathrm{opt}} \geq 0$ is its optimization tolerance, then
\begin{equation}
R_{\widehat\eta}
\leq \widehat R_{\widehat\eta}+\varepsilon
\leq \inf_{\eta\in\mathcal{A}}\widehat R_\eta+\gamma_{\mathrm{opt}}+\varepsilon
\leq \inf_{\eta\in\mathcal{A}}R_\eta+2\varepsilon+\gamma_{\mathrm{opt}}.
\end{equation}
Exact empirical minimization gives $\gamma_{\mathrm{opt}} = 0$. The proof applies on any event where the uniform estimation premise holds.

For a common fixed teacher $T$, write $R_{*}(\eta) = \|Y - \Pi_{\mathcal{S}_{\eta}} Y\|_{2}^{2}$ and $J_{\eta} = \|T - \Pi_{\mathcal{S}_{\eta}} T\|_{2}^{2}$, where $\mathcal{S}_{\eta}$ is the candidate's actual observable space. With $\delta_{T} = \|Y - T\|_{2}$, the Lipschitz property of distance gives $|\sqrt{R_{*}(\eta)} - \sqrt{J_{\eta}}| \leq \delta_{T}$. If estimates $\hat{d}_{T}(\eta)$ of $\sqrt{J_{\eta}}$ obey a uniform error $\varepsilon_{T}$, and their selector has optimization tolerance $\gamma_{\mathrm{opt}}$, then $\sqrt{R_{*}(\hat{\eta})} \leq \inf_{\eta \in \mathcal{A}} \sqrt{R_{*}(\eta)} + 2 \delta_{T} + 2 \varepsilon_{T} + \gamma_{\mathrm{opt}}$. This follows by applying the preceding selection argument to $j_{\eta} = \sqrt{J_{\eta}}$ and using the distance bound at both ends.

For multiple outputs and forecast horizons, let $f_{c, h}$ denote a target component divided by the positive training reference standard deviation $s_{c, h}$ used in Eq.~\eqref{eq:17}, and use the weighted product norm
\begin{equation}
\|f\|_\omega^2=\sum_{c,h}\omega_{c,h}\|f_{c,h}\|_2^2,
\qquad \omega_{c,h}\geq0,\qquad\sum_{c,h}\omega_{c,h}=1.
\end{equation}
After omitting zero-weight components, independent readouts at a shared configuration $\eta$ form a product Hilbert subspace. Componentwise projection gives the shared teacher residual $J_{\eta}^{\mathrm{shared}} = \sum_{c, h} \omega_{c, h} \|T_{c, h} - \Pi_{\mathcal{S}_{\eta}} T_{c, h}\|_{2}^{2}$. The risk decompositions and selection proof apply in this weighted space. A budget $\sum_{j} m_{j} \leq M_{\mathrm{tot}}$ restricts the common candidate set $\mathcal{A}$.

\section{Experimental protocols and model configurations}\label{app:B}

\subsection{Controlled population calculations}\label{app:B1}

The controlled calculations use the stationary moment solution in Eq.~\eqref{eq:7}. Classical expectations are taken under independent uniform inputs and quantum expectations are exact. The target in Eq.~\eqref{eq:12} is deterministic given the complete history. Its sine and product components are orthogonal and each has squared norm $1 / 2$. The teacher retains the complete product term $\psi_{1}(u_{t - 1}) \psi_{1}(u_{t - 5}) / \sqrt{2}$, and only the sine component is truncated.

For the delayed sine, the normalized Legendre coefficient is $\sqrt{2n + 1} (-1)^{(n - 1) / 2} j_{n}(\pi)$ for odd $n$ and zero for even $n$, where $j_{n}$ is the spherical Bessel function of the first kind. The teacher squared error is the sum of $(2n + 1) j_{n}(\pi)^{2}$ over omitted odd orders. For $p = 3, 5, 7$, this gives $4.3901166194 \times 10^{-3}$, $1.8488844975 \times 10^{-5}$, and $2.4508429952 \times 10^{-8}$, respectively. These residuals are fixed by the target and common to all coupling realizations. The degree-three teacher is used for the four-configuration decomposition and the frozen-readout history diagnostic.

At fixed $\beta$ and input law, the candidate history function span $\mathcal{D}_{\infty}$ is determined by the input moment functions and their permitted two-time products. Thus, $\varepsilon_{\mathrm{cand}}$ is common to all propagation parameters and coupling realizations at that $\beta$. The alignment error $\varepsilon_{\mathrm{align}}$ and teacher-induced readout error $\varepsilon_{\mathrm{read}}$ depend on the physical readable space. For the $20$ realizations $s = 0, \ldots, 19$, the plotted upper bound is $\bar{U}_{p} = 20^{-1} \sum_{s} (\delta_{p} + \sqrt{J_{p, s}})^{2}$. Computing each bound before averaging preserves the per-realization inequality. The history plot likewise averages $d_{L, s}$ after taking the square root for each realization.

\subsection{Mackey--Glass selection and horizon protocols}\label{app:B2}

For Mackey--Glass forecasting, each horizon $h$ is treated as a separate direct prediction task. The teacher uses a common set of Legendre-product features with maximum lag $16$, at most two active history positions, and total polynomial degree at most three. Over input ages zero through sixteen, this gives $459$ nonconstant terms. The teacher feature construction retains $294$ principal components. This dimension belongs to the teacher stage, while the six-mode quantum readout uses $90$ raw moment coordinates. All configurations at a given horizon use the same teacher construction. The readout features are standardized on training data, ridge regularization is zero, and the relative singular-value cutoff is $10^{-11}$.

The STM configuration is fixed at $(3, 1.1)$. For task alignment, $(\beta_{h}, \theta_{h})$ minimizes the normalized training teacher projection residual at each horizon using a fixed calibration coupling seed $20261013$. The search covers $\beta \in [1.02, 6]$ and $\theta \in [0.15, 3]$, using a preset grid followed by a bounded Powell search with at most $96$ function evaluations per horizon. Calibration candidates must satisfy $\|A^{1000}\|_{2} \leq 10^{-3}$. Evaluation then uses the $20$ paired static coupling seeds $0$ through $19$ with the archived disorder amplitude $0.2$. The forgetting threshold is enforced on the calibration realization, whereas each evaluation realization has its own decay scale.

The horizon extension uses common zero-based forecast origins, namely $1000, \ldots, 1654$ for training, $1800, \ldots, 1904$ for validation, and $2000, \ldots, 2904$ for testing. The corresponding counts are $655$, $105$, and $905$. Each supervised target remains within its designated partition, with the latest training target at index $1654 + 96 = 1750$. The teacher PCA and horizon-specific coefficients are fitted on the training histories and targets. Physical configurations are selected using training teacher projection scores, and all readouts are fitted to training teacher outputs and frozen before validation and test scoring. The min--max scaling is fitted on the original training prefix. The reported curves average the test NMSE of the $20$ realizations, with one standard error of the mean equal to their sample standard deviation divided by $\sqrt{20}$. Figure~\ref{fig:7} shows $h = 1, \ldots, 96$, with an inset in panel (a) enlarging $h = 1, \ldots, 20$. A separate control fixes $(1.02, 1.8)$ across horizons. When parameters are selected independently, each horizon is a separate six-mode experiment.

\subsection{Teacher-constructed passive memory}\label{app:B3}

The ETT and Weather experiments share SNAP input preparation and passive bosonic memory with the mechanism model, but use task-constructed propagation and different observation sets. For a general passive implementation, write $\hat{\boldsymbol{r}}_{t} = A \hat{\boldsymbol{r}}_{t - 1} + B \hat{\boldsymbol{a}}_{t} + E_{\mathrm{vac}} \hat{\boldsymbol{v}}_{t}$. The columns of $B$ couple fresh data ports, and $E_{\mathrm{vac}}$ couples independent fresh vacuum modes collected in $\hat{\boldsymbol{v}}_{t}$. Physical realizability requires $A A^{\dagger} + B B^{\dagger} + E_{\mathrm{vac}} E_{\mathrm{vac}}^{\dagger} = I$. The vacuum terms preserve the canonical commutators and contribute the appropriate quadrature vacuum variance. Stable moment propagation extends to this model.

For the multichannel models, a training teacher supplies linear history kernels. We project the input delay embedding and its shift dynamics onto an orthonormal basis of task-relevant delay directions. This gives a low-dimensional passive map, and vacuum ports complete the remaining defect. The reference teacher predicts each source channel's own future increments from its $336$-point history, with ridge parameter $0.1$ and all $96$ forecast horizons retained. This linear teacher defines memory directions, while the final quantum readout is fitted to the actual future increments.

Two multichannel candidates are considered. A channel bank gives each input channel its own port and memory group. A mixed-input model uses four fresh ports whose SNAP phases are dense linear combinations of all channels. The mixing weights are generated with seed $2021$, then centered and scaled as specified by the encoding construction. Each port uses either $8$ or $16$ retained modes. The actual quantum observations are the four local moments $\langle \hat{q}_{j} \rangle$, $\langle \hat{p}_{j} \rangle$, $\langle \hat{q}_{j}^{2} \rangle$, and $\langle \hat{p}_{j}^{2} \rangle$ for each mode $j$, where $\hat{q}_{j} = (\hat{r}_{j} + \hat{r}_{j}^{\dagger}) / \sqrt{2}$ and $\hat{p}_{j} = (\hat{r}_{j} - \hat{r}_{j}^{\dagger}) / (i \sqrt{2})$. The readout also includes every channel's current standardized classical value. Thus, an implementation with $m$ total modes and $C_{\mathrm{in}}$ input channels has $4m + C_{\mathrm{in}}$ non-intercept features. Its stationary readable function space includes these classical terms as well as the chosen quantum observations.

The delay-register projection gives a physical parametrization of the task-constructed memory used in the forecasting extensions. Let $D_{\mathrm{lag}}$ be the length of a register that stores one input at each delay, let $T_{\mathrm{sh}} \in \mathbb{R}^{D_{\mathrm{lag}} \times D_{\mathrm{lag}}}$ shift each coordinate to the next older lag, and let $e_{0}$ be the vector injecting the newest input. They satisfy $T_{\mathrm{sh}} T_{\mathrm{sh}}^{\top} + e_{0} e_{0}^{\top} = I_{D_{\mathrm{lag}}}$. Let $V_{\mathrm{lag}} \in \mathbb{C}^{D_{\mathrm{lag}} \times m}$ contain orthonormal retained directions selected from the training-kernel observability modes, so that $V_{\mathrm{lag}}^{\dagger} V_{\mathrm{lag}} = I_{m}$. The compressed map is
\begin{equation}
A=V_{\mathrm{lag}}^\dagger T_{\mathrm{sh}}V_{\mathrm{lag}},
 \qquad b=V_{\mathrm{lag}}^\dagger e_0,
 \qquad
 E_{\mathrm{vac}}E_{\mathrm{vac}}^\dagger=I_m-AA^\dagger-bb^\dagger.
\end{equation}
The matrix $V_{\mathrm{lag}} V_{\mathrm{lag}}^{\dagger}$ is an orthogonal projector, so $A A^{\dagger} + b b^{\dagger} \leq I_{m}$. The remaining positive semidefinite defect can therefore be factored into vacuum couplings $E_{\mathrm{vac}}$. This finite-register compression is stable. Indeed, a unit-modulus eigenvalue of $A$ would saturate both norm inequalities in $\|V_{\mathrm{lag}}^{\dagger} T_{\mathrm{sh}} V_{\mathrm{lag}} x\|_{2} \leq \|T_{\mathrm{sh}} V_{\mathrm{lag}} x\|_{2} \leq \|V_{\mathrm{lag}} x\|_{2}$ for its nonzero eigenvector $x$. The first equality would place $T_{\mathrm{sh}} V_{\mathrm{lag}} x$ in the retained subspace, making $V_{\mathrm{lag}} x$ an eigenvector of $T_{\mathrm{sh}}$ with the same unit-modulus eigenvalue. This contradicts the nilpotence of the finite shift, and hence $\operatorname{spr}(A) < 1$. The vacuum modes are fresh at every step and are not counted as persistent memory. The input controllability Gramian can be smaller than $I_{m}$ because these additional ports also remove information.

The model uses an eight-level input state. When evaluating quadrature powers, the input operator representation is padded to dimension $12$, so the intermediate ladder operations are evaluated without an artificial boundary at the highest occupied level. This numerical padding does not change the occupied input support. The persistent bosonic evolution is evaluated through exact moments. The multichannel models read out the memory after the current input is injected, so their response includes age zero.

\subsection{Forecasting data, fitting, and model configurations}\label{app:B4}\label{app:B5}

\textbf{Chronological splits and preprocessing.} Table~\ref{tab:2} gives the chronological segment lengths. All origins have a complete $96$-step target inside the segment being scored. For training length $N_{\mathrm{tr}}$, validation length $N_{\mathrm{va}}$, and test length $N_{\mathrm{te}}$, the zero-based origin sets are $1000, \ldots, N_{\mathrm{tr}} - 97$ for training, $N_{\mathrm{tr}} - 1, \ldots, N_{\mathrm{tr}} + N_{\mathrm{va}} - 97$ for validation, and $N_{\mathrm{tr}} + N_{\mathrm{va}} - 1, \ldots, N_{\mathrm{tr}} + N_{\mathrm{va}} + N_{\mathrm{te}} - 97$ for testing. The state is propagated from the available chronological prefix. The first $1000$ origins are excluded from supervised training, while validation and test continue to use earlier observed history.

\begin{table}[tbp]
\centering
\caption{\textbf{Multichannel segment lengths and common forecast origins.} All counts are in data rows. The nominal intervals are one hour for ETTh, fifteen minutes for ETTm, and ten minutes for Weather.}
\label{tab:2}
\begingroup
\small
\setlength{\tabcolsep}{4pt}
\renewcommand{\arraystretch}{1.22}
\begin{tabularx}{\linewidth}{@{}l*{6}{>{\centering\arraybackslash}X}@{}}
\toprule
\textbf{Dataset} & \textbf{Segment train} & \textbf{Segment validation} & \textbf{Segment test} &
\textbf{Origins train} & \textbf{Origins validation} & \textbf{Origins test} \\
\midrule
ETTh1, ETTh2 & $8640$ & $2880$ & $2880$ & $7544$ & $2785$ & $2785$ \\
ETTm1, ETTm2 & $34560$ & $11520$ & $11520$ & $33464$ & $11425$ & $11425$ \\
Weather & $36887$ & $5270$ & $10539$ & $35791$ & $5175$ & $10444$ \\
\bottomrule
\end{tabularx}
\endgroup
\end{table}

Encoding uses each channel's training minimum and maximum to map the training range to $[-1, 1]$. Subsequent inputs are not clipped and may fall outside that interval. Targets and current classical input features are standardized using complete training-segment means and population standard deviations. The saved data are used without imputing or clipping values, including the recorded Weather wind-speed outlier. Forecast steps are indexed by data rows. For a constant output channel, the implementation reports raw MSE and excludes its undefined NMSE from the macro average. The evaluated channels in the reported main comparison have positive training variance.

\textbf{Multichannel fitting and resource counts.} Each final head predicts the standardized increment from the current value to each future horizon. The current value is added back before restoring physical units. Feature centering and scaling are estimated on supervised training origins. For each candidate ridge value, the fitted $96$-horizon coefficient matrix and intercept are projected onto each prescribed horizon basis. The ridge grid is $\{0.001, 0.1, 10, 100\}$, and the horizon-rank grid is $\{8, 16, 96\}$. A rank of $96$ leaves the horizon axis unrestricted. This rank acts on the forecast-output axis. The ridge value and horizon rank are selected independently for every output, and the physical candidate is selected by the output-averaged validation NMSE. The saved selection is fixed before test scoring. The five reported $M$ models use seed $2021$ and no calendar head, with resource counts listed in Table~\ref{tab:3}.

\begin{table}[tbp]
\centering
\caption{\textbf{Selected multichannel implementations.} The quantum feature count is $4m$, and current classical input values add $C_{\mathrm{in}}$ coordinates. The intercept is excluded from the last column. All listed modes are persistent memory modes.}
\label{tab:3}
\begingroup
\small
\setlength{\tabcolsep}{4pt}
\renewcommand{\arraystretch}{1.22}
\begin{tabularx}{\linewidth}{@{}l*{6}{>{\centering\arraybackslash}X}@{}}
\toprule
\textbf{Dataset} & \textbf{Structure} & \textbf{Ports} & \textbf{Modes/port} & \textbf{Total $m$} &
\textbf{Quantum features} & \textbf{Total features} \\
\midrule
ETTh1 & Channel bank & $7$ & $8$ & $56$ & $224$ & $231$ \\
ETTh2 & Mixed input & $4$ & $8$ & $32$ & $128$ & $135$ \\
ETTm1 & Channel bank & $7$ & $16$ & $112$ & $448$ & $455$ \\
ETTm2 & Mixed input & $4$ & $8$ & $32$ & $128$ & $135$ \\
Weather & Mixed input & $4$ & $8$ & $32$ & $128$ & $148$ \\
\bottomrule
\end{tabularx}
\endgroup
\end{table}

In the ETT output order HUFL, HULL, MUFL, MULL, LUFL, LULL, OT, the selected ridge parameters are $(0.1, 0.1, 0.1, 0.1, 100, 10, 0.1)$ for ETTh1, $(10, 10, 10, 10, 10, 10, 0.1)$ for ETTh2, $(0.1, 0.1, 0.1, 0.1, 10, 10, 10)$ for ETTm1, and $(0.1, 10, 0.001, 10, 0.1, 10, 0.1)$ for ETTm2. All these outputs use horizon rank $96$, except ETTh2 HULL, which uses rank $16$. The complete per-output parameter records, including Weather, are provided with the manuscript source package.

\textbf{Single-output references.} The single-output values in the forecasting comparison correspond to the selections in Table~\ref{tab:4}. Each uses a $32$-mode core, and the corresponding branches include the reference-memory and retention components. The table reports the core observation order. Delta branches use local quadrature moments through order four under two passive output settings, and level branches use moments through order two under one setting. Both quadratures are evaluated in each setting. Each $S$ construction therefore has its own observation set and mode count.

\begin{table}[tbp]
\centering
\caption{\textbf{Single-output model identity and teacher settings.} History length is measured in data rows. The moment order describes the core quantum features. The listed versions exclude calendar inputs.}
\label{tab:4}
\begingroup
\small
\setlength{\tabcolsep}{4pt}
\renewcommand{\arraystretch}{1.22}
\begin{tabularx}{\linewidth}{@{}l>{\raggedright\arraybackslash}p{0.29\linewidth}XXX@{}}
\toprule
\textbf{Dataset} & \textbf{Selected branch} & \textbf{Teacher history} & \textbf{Teacher ridge} &
\textbf{Core order} \\
\midrule
ETTh1 & Delta with retention & $336$ & $0.1$ & $4$ \\
ETTh2 & Delta with reference memory & $672$ & $1$ & $4$ \\
ETTm1 & Level core & $672$ & $0.1$ & $2$ \\
ETTm2 & Delta with reference memory & $672$ & $0.01$ & $4$ \\
Weather & Level core & $168$ & $0.01$ & $2$ \\
\bottomrule
\end{tabularx}
\endgroup
\end{table}

\subsection{Training protocols and hyperparameters for classical baselines}\label{app:B6}

The PatchTST, DLinear, and Autoformer references are trained in this work, not copied from the original publications. All three use the same data splits and the same input length ($L_{\mathrm{in}} = 336$ rows) as the corresponding SNAP QRC configuration. Table~\ref{tab:baselines} lists the configurations. PatchTST and DLinear follow their published settings, and Autoformer uses the configuration of the original long-horizon benchmarks.

\textbf{Horizon protocol.} Each classical baseline is trained separately for each prediction length, following the standard benchmarking convention. For $h = 96$, this means one model with a $96$-step output. SNAP QRC uses a different protocol in which a single fit produces all $96$ horizon outputs through independent per-horizon heads whose ridge and horizon rank are selected on validation data.

\textbf{Normalization and denominator.} All inputs and targets are standardized with the training-segment mean and population standard deviation, both for SNAP QRC and for the classical references. The reported NMSE divides by the same training-segment variance in every row, so all methods share a single denominator.

\begin{table}[p]
\centering
\caption{\textbf{Archived configurations of the classical baselines used for the $h=1$ and $h=96$ comparisons.} Each prediction length is trained separately. Blank entries indicate settings that are not applicable to the implemented architecture.}
\label{tab:baselines}
\begingroup
\small
\setlength{\tabcolsep}{5pt}
\renewcommand{\arraystretch}{1.16}
\begin{tabularx}{\linewidth}{@{}>{\raggedright\arraybackslash}p{0.29\linewidth}*{3}{>{\raggedright\arraybackslash}X}@{}}
\toprule
\textbf{Setting} & \textbf{PatchTST} & \textbf{DLinear} & \textbf{Autoformer} \\
\midrule
Input length & $336$ & $336$ & $336$ \\
Architecture & $3$ encoder layers & $2$ parallel linear branches & $2$ encoder, $1$ decoder layers \\
Embedding dimension & $16$ &  & $512$ \\
Attention heads & $4$ &  & $8$ \\
Feedforward dimension & $128$ &  & $2048$ \\
Activation & GELU &  & GELU \\
Model dropout & $0.3$ &  & $0.05$ \\
Forecasting-head dropout & $0$ &  &  \\
Patch length / stride & $16/8$ &  &  \\
Moving-average window &  & $25$ & $25$ \\
Decoder context length &  &  & $48$ \\
Autocorrelation factor &  &  & $3$ \\
RevIN & Enabled &  &  \\
\midrule
Batch size & $128$ & $128$ & $32$ or $128$\textsuperscript{a} \\
Maximum epochs & $100$ & $100$ & $10$ \\
Optimizer & Adam & Adam & Adam \\
Training loss & MSE & MSE & MSE \\
Configured learning rate & $10^{-4}$ & $10^{-4}$ & $10^{-4}$ \\
Learning-rate schedule & Exponential decay\textsuperscript{b} & Exponential decay\textsuperscript{b} & Exponential decay\textsuperscript{b} \\
Early-stopping patience & $100$ epochs & $100$ epochs & $100$ epochs \\
Checkpoint selection & Lowest validation loss & Lowest validation loss & Lowest validation loss \\
Random seed & $2021$ & $2021$ & $2021$ \\
\bottomrule
\end{tabularx}
\par\smallskip
\begin{minipage}{\linewidth}
\footnotesize
\textsuperscript{a} Autoformer uses batch size $128$ for ETTh1 single-channel $h=96$ and ETTh2 single-channel and multichannel $h=96$, whereas the remaining reported configurations use $32$. These values follow the final run recorded for each configuration.\\[3pt]
\textsuperscript{b} The learning rate is set at the end of epoch $e$ to $10^{-4}\,0.9^{\max(e-3,0)}$.\\[3pt]
PatchTST uses end padding, a shared flattened prediction head, and RevIN with affine parameters disabled and mean subtraction. Its decomposition option is disabled. The forecasting head uses zero dropout. DLinear applies separate $336\!\to\!h$ linear maps to the seasonal and trend components, with weights shared across channels. Autoformer uses continuous time-feature embedding and has no distilling stage in the archived implementation. All methods use training-segment standardization, and RevIN is an additional PatchTST operation.
\end{minipage}
\endgroup
\end{table}

\textbf{The $h=1$ behavior of Autoformer.} The Autoformer $h=1$ errors in Table~\ref{tab:1}(a) exceed those of Persistence by roughly an order of magnitude. This is a consequence of applying the published long-horizon configuration at a horizon length for which it was not designed, not a mismatch in the evaluation protocol. On ETTh1, the training loss decreases monotonically over the ten training epochs and the validation loss is still decreasing at the final epoch ($0.0645$), so the model is undertrained in absolute terms, while its $512$-dimensional representation is far larger than a one-step target requires. Evaluated on the $2785$ common test forecast origins used in Table~\ref{tab:1}(a), the test NMSE is $0.07691$, roughly seventeen times the PatchTST value. We report the published configuration unchanged rather than tuning it per horizon, so that the comparison reflects the baselines as they are normally deployed.

\end{document}